\documentclass{ieeeaccess}
\usepackage{cite}
\usepackage{amsmath,amssymb,amsfonts}
\usepackage{algorithmic}
\usepackage{graphicx}
\usepackage{textcomp}
\usepackage{multirow}
\usepackage{multicol}
\usepackage{setspace}
\usepackage{fancyhdr}
\usepackage[dvipsnames]{xcolor}
\usepackage{graphicx} %Loading the package
\usepackage{lipsum}
\graphicspath{ {Figures/} }
\usepackage{array}
\usepackage{float} %float figures tables & algorithm
\usepackage{dblfloatfix}
\usepackage{amssymb,amsmath,amsfonts,mathtools} %mathematical tools
\usepackage{tabto}
\usepackage{subcaption}
\usepackage{soul}

\usepackage[switch, pagewise]{lineno}
\usepackage{url}
\usepackage[hidelinks]{hyperref}
\usepackage{nicefrac}
\usepackage{rotating}
\usepackage{lscape}
\usepackage{ragged2e}

\usepackage{caption,setspace}
\usepackage{color,soul}
\newcommand{\comment}[1]{}

\def\BibTeX{{\rm B\kern-.05em{\sc i\kern-.025em b}\kern-.08em
    T\kern-.1667em\lower.7ex\hbox{E}\kern-.125emX}}
\begin{document}
 \history{Date of publication xxxx 00, 0000, date of current version xxxx 00, 0000.}
 \doi{xx.xxxx/ACCESS.xxxx.DOI}

\title{Transmittance Multispectral Imaging for Reheated Coconut Oil Differentiation}

\author{
\uppercase{D.~Y.~L.~Ranasinghe}\authorrefmark{1},
\uppercase{H.~K.~Weerasooriya}\authorrefmark{1},
\uppercase{S.~Herath}\authorrefmark{2},
\uppercase{M.~P.~Bandara Ekanayake}\authorrefmark{1},\IEEEmembership{Senior Member, IEEE},
\uppercase{H.~M.~V.~R.~Herath}\authorrefmark{1},\IEEEmembership{Senior Member, IEEE},
\uppercase{G.~M.~R.~I.~Godaliyadda}\authorrefmark{1},\IEEEmembership{Senior Member, IEEE},
\uppercase{Terrence Madhujith}\authorrefmark{3}
}

\address[1]{Department of Electrical and Electronic Engineering, Faculty of Engineering, University of Peradeniya, Peradeniya, Sri Lanka (20400)}
\address[2]{Department of Electrical and Computer Engineering, University of Maryland, College Park, MD, USA (21250)}
\address[3]{Department of Food Science and Technology, Faculty of Agriculture, University of Peradeniya, Peradeniya, Sri Lanka (20400)}

\corresp{Corresponding author: D.Y.L Ranasinghe (e-mail: e14273@eng.pdn.ac.lk).}

\tfootnote{This work was supported by University of Peradeniya, Sri Lanka research grant (No: URG/2017/26/E).}

\markboth
{Ranasinghe \headeretal: Transmittance multispectral imaging for reheated coconut oil differentiation}
{Ranasinghe \headeretal: Transmittance multispectral imaging for reheated coconut oil differentiation}

\begin{abstract}
Oil reheating has a significant impact on global health due to its extensive consumption, especially in South Asia, and severe health risks.
Nevertheless, food image analysis using multispectral imaging systems(MISs) has not been applied to oil reheating analysis despite their vast application in rapid food quality screening.
To that end, the paper discusses the application of a low-cost MSI to estimate the `reheat cycle count classes' (number of times an oil sample is recursively heated) and identify `critical classes' at which substantial changes in the oil sample have materialized.
Firstly, the reheat cycle count class is estimated with Bhattacharyya distance between the reheated and a pure oil sample as the input. The classification was performed using a support vector machine classifier that resulted in an accuracy of 83.34\,\% for reheat cycle count identification.
Subsequently, an unsupervised clustering procedure was introduced using a modified spectral clustering (SC) algorithm to distinguish critical classes under reheating. In addition, laboratory experiments were performed to ascertain the ramifications of the reheating process with a chemical analysis. The chemical analysis of the coconut oil samples used in the experiment coincided with the chemical analysis results and was statistically significant ($p < 0.05$).
Accordingly, the proposed work closes the gap for using multispectral imaging for oil reheating and proposes a novel algorithm for unsupervised detection of critical property changes in the oil. Hence, the proposed research work is significant in its practical implications, contribution to food image analysis, and unsupervised classification mechanisms.
\end{abstract}


\titlepgskip=-15pt

\maketitle

%\onecolumn
%\linenumbers

\section{Introduction}
\label{section:introduction}
The quality of food consumed plays a pivotal role in assuring the health of a society. Therefore, it is of paramount importance to continuously monitor the quality of food. Rapid screening of food and beverages, including edible oils,  has become a key focus among scientists and industrialists because contamination and adulteration of food compromise the quality of food \cite{lim2020pattern, popa2020rapid, subhi2019vision}. In this context, the need for accurate, fast, non-destructive, and economical methods to assure the standard of food are of a timely need.

In recent years, numerous sensor-based techniques and systems have been proposed and implemented to learn individual dietary and energy intake \cite{farooq2016novel,prioleau2017unobtrusive,farooq2016segmentation}, identify unique food items in meals \cite{lopez2012automatic,he2013food}, and monitor food consumption characteristics such as food microstructure \cite{doulah2017meal}. Among the sensor-based food assessment, techniques based on image analysis are a prominent research avenue for rapid screening of food and beverages. Red-Green-Blue (RGB) colour photography, spectroscopy, and spectral imaging (SI) could be food image analyses. In RGB photography, recorded images are reconstructed by mixing available ground-truths for the three spectral regions in various proportions. This technique has been used in a wide range of applications: detection of skin defects in citrus fruits \cite{lopez2010automatic}, development of sorting and grading mechanisms \cite{al2011computer, zhang2015computer}, and dietary assessment via food image analysis with deep learning \cite{jiang2020deepfood}. Although RGB photography has its merits in food analyses, the technique has spectral limitations when analyzing items responsive to UV-region, such as oil and vinegar. Besides that, spectroscopy and SI techniques are superior to RGB photography at sensing and differentiating such items as these methods include signatures from outside the visible region. Besides that, spectroscopy and SI provide better feature discrimination power with greater spectral bands than RGB photography.

In particular, spectroscopy methods are useful in deriving elaborative quality parameters  \cite{jamwal2020application,nunes2014vibrational, fengou2020estimation} using spectral characteristics. Especially, spectroscopy techniques such as Fourier Transform Infra-Red (FTIR) \cite{jamwal2020rapid}, and Raman spectroscopy \cite{li2019identification} are regularly used in the compositional analysis of oil and agricultural produce \cite{manaf2007analysis}. Though both the above spectroscopy techniques have disparate operating spectral regions, both methods can derive the respective spectrum of the specimen within minutes. Nonetheless, spectroscopy techniques are contingent upon established reference spectra similar to RGB photography for calibration and baseline removal. Hence, without accurate references, the potential of this method is severely hindered. Besides that, spectroscopy equipment is complex, expensive, and restricted to a laboratory environment rather than for field use compared to regular RGB images. Moreover, the method requires professional knowledge and experience to operate the instrument and analyze the measurements.

SI is a cost-effective alternative to spectroscopy methods which could be described as an extension of RGB photography into near-infrared (NIR) and ultraviolet (UV) regions \cite{khan2018modern}. Similar to spectroscopy, SI could be used to analyze both solids \cite{li2020pickled} and liquids \cite{qin2007measurement}, as well as opaque and translucent materials. SI could be categorized as hyperspectral imaging systems (HISs) and multispectral imaging systems (MISs) depending on the spectral discriminating power each system offers. Usually, HISs have more discrete wavelengths with a narrow spectral resolution, and the monochrome spectral images are acquired with a diffraction mechanism. Whereas MISs only utilize a few selected wavelengths ranging from NIR to UV and use either optical filters with ambient light or separate light sources for each spectral band along with a monochrome camera for image acquisition. In the literature, hyperspectral imagery has been used to extract information about bruises in peaches \cite{pan2016detection}, and strawberries \cite{nagata2006bruise} with significant accuracies. Moreover, HISs have been used to analyze skim and nonfat milk powder \cite{harnly2014characterization} and to predict microbial spoilage of bakery products \cite{saleem2020prediction} which demonstrate the applicability of HISs at various levels in the food chain. Although HISs offer superior spectral features to MISs, it demands complex systems to control and standardize the image acquisition process.
Additionally, it costs about USD 8000 \cite{stuart2020low} to build a low-cost HIS when the average GDP per capita in South Asian countries, where oil reheating is a more prevalent issue\cite{kakde2017urbanized, arachchige2021potential}, is around USD 2000 \cite{gdp2020worldbank}. It can be argued that the HIS is not suitable for on spot testing instead of the proposed imaging setup considering the cost and hardware complexity of a HIS compared to an MIS. Besides that, MISs could be used as an on-spot primary screening mechanism to separate aberrant samples, which then can be analyzed extensively with HIS, spectroscopy or chemical methods.
Nevertheless, the development of an MIS imposes an optimization problem for the band selection in a cost-effective manner. Since MISs integrate both imaging and spectroscopy techniques, MISs offer advantages such as minimal sample preparation, non-destructive examination, and fast-acquisition times over chemical analyses \cite{gowen2007hyperspectral, rahi2018spectroscopy, wang2007spectral}. These practical advantages, along with the development of computational imaging, have fomented exploring MISs as a viable option for various applications in food image analysis \cite{brosnan2004improving,cheng2015rapid,elmasry2008early} amongst researchers. Furthermore, MISs are simpler and less expensive to build and could be used as a portable device, as evidenced by the developed MIS in the proposed work instead of HISs.

Reheating is another form of food preservation among injecting chemicals to ripen, preserving perishable goods, but done for cost mitigation. In particular, this practice is commonplace in South Asian countries \cite{kakde2017urbanized, arachchige2021potential}, especially at households and restaurants. Though reheating and reusing oil has its financial incentive, such corruption is pernicious to health. 
Consequently, the ramifications of oil reheating are significant because the degree of harm from consumption and the total population of consumers is high. Hence, it is a clear requirement to flag if such appreciable change has occurred and is detected in the oil. Besides that, there are no rapid screening techniques to identify such changes due to oil reheating, even though they happen vastly in fast food chains.
However, several assessment techniques were proposed in the past few decades to estimate various quality parameters of edible oils with MISs under different degrading processes like adulteration and toxin formation during the manufacturing stage. At the same time, only a limited amount of work has been carried out to assess the effects of reheating and reusing the oil. Since repetitive use of coconut oil is expected in food-service establishments and households as a measure to cut down the cost \cite{bhardwaj2016effect}, coconut oil was used to obtain experimental results in the proposed work. Though reheating is a continuous process, the number of times oil is reheated is a discrete variable. Therefore, contamination due to reheating can happen at different levels and can have significant alterations at different reheat cycles. Hence, it is a clear requirement to flag if such appreciable change has occurred and been detected. Since such demarcations can be done for the same oil under different reheat cycles, the change detection must be unsupervised. It is necessary to treat this as a specific novel unsupervised research problem in food image analysis. 
Moreover, the chemical and thermophysical properties are altered during reheat \cite{bhuiyan2016determination} and these physicochemical changes compromise the safety of the oil, thus potentially making fried foods unsafe for consumption \cite{guillaume2018evaluation}. Many secondary oxidative products such as carbonyls, organic acids, hydrocarbons, and polymerized compounds are generated during repeated heating. Furthermore, the production of trans-fatty acids raises low-density lipoprotein (LDL) and total cholesterol while decreasing high-density lipoprotein (HDL) cholesterol. Consequently, often use of impure coconut oil leads to an increased risk of cardiovascular diseases \cite{khera2011cholesterol, hernaez2019role}, carcinogenesis, and other non-communicable diseases \cite{hamsi2015effect}. Though it is patent that the formation of secondary products is unhealthy, it is difficult to determine the healthiness and unhealthiness of contaminated samples; instead, food analyses are more concerned about the relative change in the chemical composition of the sample; specifically, the oxidative stability \cite{ghazali2009oxidative, karimi2017impact, seneviratne2016phenolic, melo2019first} since it involves significant health risks.

\begin{figure*}[b!]
\centering
\begin{subfigure}[t]{0.95\columnwidth}
    \includegraphics[width=0.95\columnwidth]{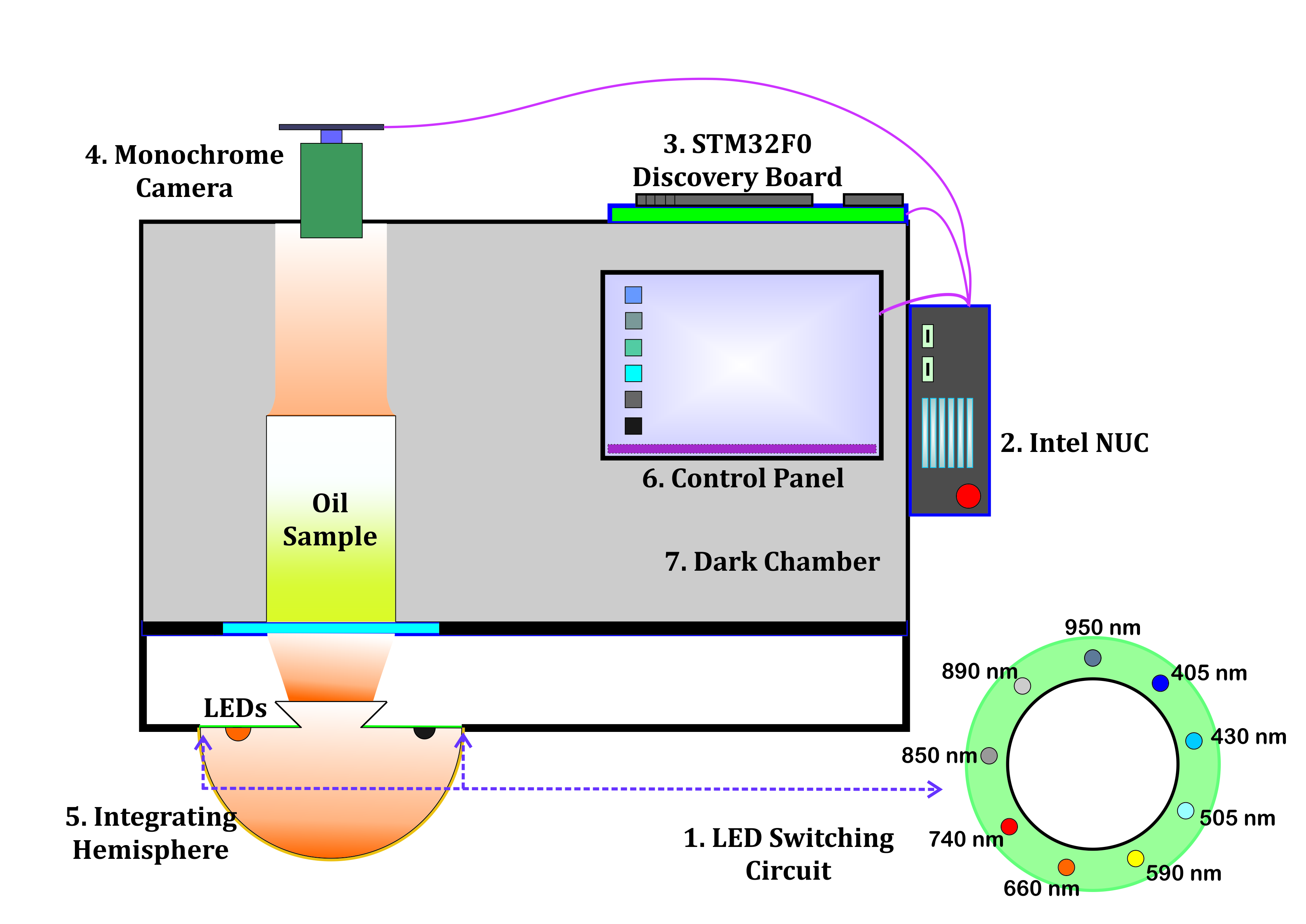}
    \captionsetup{justification=centering}
    \caption{}
    \label{figure: msi setup}
\end{subfigure}
~
\begin{subfigure}[t]{0.95\columnwidth}
    \includegraphics[width=0.95\columnwidth]{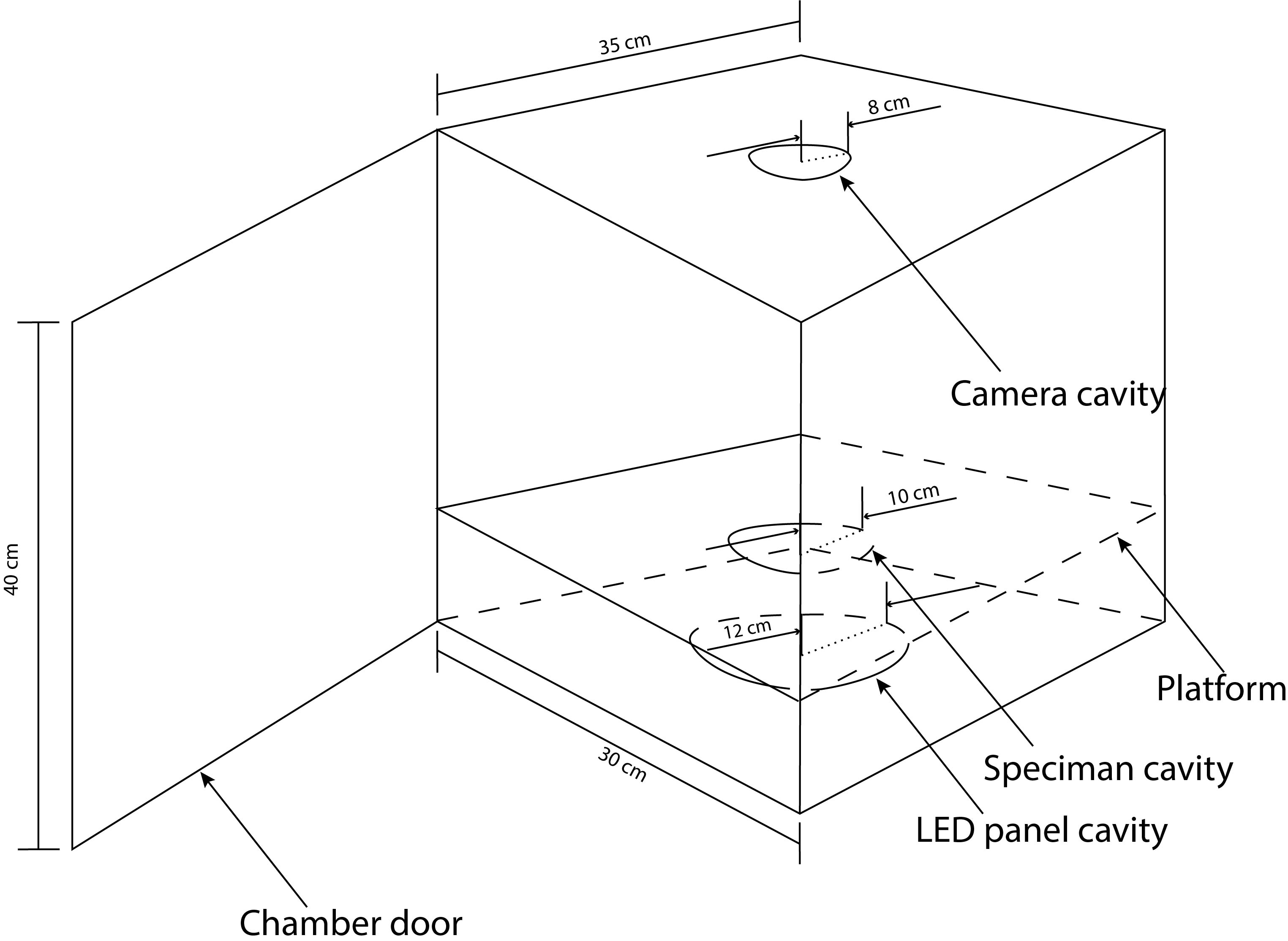}
    \captionsetup{justification=centering}
    \caption{}
    \label{figure: msi design}
\end{subfigure}
\caption{(a) Schematic diagram and (b) design (obscure edges and radii are depicted in dashed lines and dotted lines, respectively) of the in-house developed transmittance based MIS}
\end{figure*}

This work discusses a novel application for MISs with coconut oil as the case study. As articulated above, spectral imaging has not been applied to study reheating in food quality analysis, let alone coconut oil. Hence, the application of multispectral imaging to oil reheating is a novel contribution to food image analysis. 
Then, the proposed work discusses two analytical methods: reheat cycle count classification and significant alteration detection. The former is similar to algorithms proposed in adulteration for constructing a functional relationship. However, this functional relationship is used for classification for the class estimation problem, unlike in oil adulteration studies. However, the latter proposes a novel usage for oil contamination detection where the primary agenda is to find at what point significant property change occurs. We are proposing using unsupervised classification to identify appreciable changes in spectral properties. Further, we propose a novel mechanism to determine the number of clusters with spectral clustering.
The proposed application estimates the reheat cycle count class (number of times a given oil sample has been reheated) compared to a fresh batch of the oil and can be used to recognize reheat cycle count classes with significant chemical alterations to the oil during the reheating process. For brevity, the reheat cycle count classes where these dramatic changes in the composition occur will be referred to as `critical reheat cycle count classes'. The method proposed in this work utilizes the transmittance of light through the specimen to produce a transmittance spectrum of the oil sample. Then, the algorithm developed to estimate the reheat cycle count classes first incorporates Fisher Discriminant Analysis (FDA) for dimension reduction. These selected features were used to calculate Bhattacharyya distance, which the Support Vector Machine used to estimate the reheat cycle count. Next, spectral clustering was used to build an algorithm to distinguish the reheat cycle count classes with significant spectral property changes. The outcome of the grouping is the formation of umbrella clusters of reheat cycle count classes with incremental changes in spectral properties due to reheating. As for reheating causes changes in chemical properties, cluster boundaries could perhaps imply critical changes in composition, indicating health risks. 

The proposed work contributes to multispectral imaging under the food image analysis research, and we are proposing two analytical methods for oil reheating. 
Firstly, to determine the reheat level count class and secondly to detect significant chemical property changes of the oil. Since the reheat level count class determination is a classification task, we have compared different classifiers using Bhattacharyya distance as the input feature. Finding a suitable classifier under the proposed feature selection method is one of the technical contributions of this work.
Secondly, we proposed a novel mechanism to detect marked changes in the oil, in particular oxidative stability of the oil. For that, we are proposing a novel mechanism to determine the number of clusters using spectral clustering to select the optimal number of clusters. Furthermore, we compared the algorithm results with the chemical analysis because multispectral imaging has not been applied in analyzing property changes.
The contributions of the work can be listed as follows:
(a) An in-house developed portable transmittance-based MIS. 
(b) A rapid and cost-effective screening algorithm for reheated coconut oil.
(c) A methodology that estimates the number of times the oil has been reheated.
(d) The proposed methodology can identify the reheat cycle counts at which a notable change in the oil has happened via monitoring spectral properties indicative of significant chemical and thermophysical property changes.

The paper is organized as follows. First, a summary of related works in MISs in food image analysis is included in section \ref{section: related work}. Followingly, specifications of the MIS, experimental procedure, image pre-processing, and algorithm development are expounded in section \ref{section: materials and methods}. Lastly, the results of the developed algorithms and the chemical analysis for the proposed application are available in section \ref{section: results} along with the evaluation in section \ref{section: discussion}.

\begin{table*}[b!]
\centering

\caption{Specifications of the LED panel of the MIS}
\resizebox{\textwidth}{!}{
\begin{tabular}{c c c c c}
\hline \hline \\[-0.7ex]
LED Number 	& Typical peak wavelength (nm)	& Typical emitting spectral band (nm) & Typical half power bandwidth (nm)\\[1ex]
\hline
\\[-1ex]
1	& 405   &	375 -- 435	& 20\\[0.4ex] 
2	& 430   &	375 -- 475	& 20\\[0.4ex] 
3	& 500	&   450 -- 550	& 30\\[0.4ex] 
4	& 610	&	520 -- 620	& 10\\[0.4ex] 
5	& 660	&	630 -- 685	& 20\\[0.4ex] 
6	& 740	&	670 -- 770	& 20\\[0.4ex] 
7	& 850	&	770 -- 900	& 30\\[0.4ex] 
8	& 890	&	830 -- 970	& 40\\[0.4ex] 
9	& 950	&	900 -- 1000	& 50\\[0.4ex] 
\hline
\hline
\label{table: led panel specifications}
\end{tabular}}
\end{table*}

\section{Related work}
\label{section: related work}
Multispectral imaging systems have been used in applications from various fields such as agriculture \cite{li2020analysis}, microbiology \cite{fengou2020estimation}, entomology \cite{fennell2018method}, etc. In specific, adulteration and defects of agricultural produce have been analyzed using MIS for turmeric \cite{bandara2020validation}, fruits \cite{hashim2018evaluation, ariana2006integrating, nguyen2020precise}, beef \cite{ropodi2015multispectral}, and rice \cite{liu2014nondestructive}, as well as in packed foods \cite{senni2016multispectral}, and tomato paste \cite{liu2017potential}. Besides that, different imaging systems have been proposed in the literature with applications. For example, the growth of mould on food has been examined \cite{ebrahimi2015quantitative} using the VideometerLab spectral imaging instrument with 18 spectral bands. Also, an MIS operating in the spectral range of 405–970\,nm consisting of 19 different spectral bands has been used to determine the aerobic plate count of cooked pork sausages \cite{ma2014multispectral}. In addition, SI has been used for crop monitoring, and vegetation index calculation \cite{de2018low} and to assess the health status of crops \cite{honrado2017uav}. In \cite{raju2020detection}, an MIS  has been developed to detect oil-containing dressing on salad leaves to assess the energy intake. Furthermore, SI has been used to analyze coconut oil adulteration with palm oil \cite{weerasooriya2020transmittance}, mustard oil \cite{jamwal2020utilizing}. The work proposed in \cite{palananda2021turbidity} presents the determination of turbidity in coconut oil with food image analysis.

In most work found in the literature for studies on edible oil, the prime focus has been natural degradation, adulteration, and defect identification. However, processes such as oil reheating and associated changes have not been under scrutiny with SI as much compared to other debasing processes: oil adulteration and toxin formation, even though the chemical changes from reheating are equally deleterious. In particular, oil adulteration is the most extensively studied contamination method in the food research community using chemical analyses \cite{frankel2010chemistry,jabeur2014detection,chen2018detection} and spectroscopy methods \cite{rohman2009analysis,quinones2013detection,baeten1996detection,rohman2011use,jamwal2020rapid}. However, the application of spectral imaging to study oil adulteration has been limited, let alone coconut oil, even though spectral imaging has been used to detect adulterants in consumables such as wheat flour in turmeric \cite{bandara2020validation}, limestone powder in tapioca starch \cite{khamsopha2021utilizing}, horsemeat in beef \cite{ropodi2017multispectral}, sucrose in tomato paste \cite{liu2017potential}.

In \cite{herath2020quantitative,weerasooriya2020transmittance} for oil adulteration studies, the focus has been on developing a functional relationship since the requirement is to calculate the growth of a parameter. However, in the adulteration analysis using MSIs, the change detection in chemical properties has not been studied as opposed to the proposed work for reheating analysis. Unlike the adulteration or reheat cycle count class estimation, which uses supervised learning, the discrimination of appreciable chemical property changes has been performed with unsupervised learning. Then, toxin formation detection in oil has been studied in \cite{kumar2021assessing,chen2021solid} using mass spectrometry and FTIR spectroscopy, respectively. Though multispectral imagery has not been applied in toxin formation in oil so far to the best of our knowledge, considering the parallelity of spectroscopy and spectral imaging, it should be possible to develop an algorithm to detect the presence aflatoxin in oil.

\section{Materials and methods}
\label{section: materials and methods}

\subsection{Development of the multispectral imaging system}
\label{development of the MIS}
Most MISs found in the literature have been designed to derive the reflectance spectrum of the specimen, often of opaque and solid materials. However, this configuration is futile with translucent liquids such as oils and vinegar because only a fraction of light is reflected while most light is transmitted through the sample. To that end, the MIS developed for this application was configured as depicted in Fig. \ref{figure: msi setup} to measure the transmittance spectrum of liquids which is an adaptation of the reflectance-based setup proposed in \cite{goel2015hypercam, prabhath2019multispectral}. The optical excitation for the sample is generated using a dedicated illumination panel comprised of narrow-band LEDs corresponding to nine spectral bands from 375\,nm to 1000\,nm with specifications as given in Table \ref{table: led panel specifications}, accompanied by an illumination panel made of~~  Aluminum~~  with a 130\,mm inner~~ diameter~~ to~~ provide~~ better~~ illumination~~ for~~ the 

\begin{landscape}
    \begin{table}[h]
    \centering
    
    \caption{Details of the components for the MIS}
        \begin{tabular}{>{\raggedright}p{25mm}>{\noindent\justifying}p{70mm}>{\raggedright}p{50mm}>{\noindent\justifying}p{40mm}>{\arraybackslash}p{25mm}}
        \hline \hline \\[-0.7ex]
        Item &Detailed Description &Manufacturer Product Number &Manufacturer &Headquaters\\[1ex]
        \hline\\[-1ex]
        IC - 1
        &Linear Voltage Regulator IC Positive Fixed 1 Output 300mA 8-MSOP
        &\href{https://www.digikey.com.au/products/en?keywords=ADP3333ARMZ-3.3-R7CT-ND\%20}{ADP3333ARMZ-3.3-R7}
        &Analog Devices Inc.
        &United States\\[4ex]
        IC - 2
        &Linear Voltage Regulator IC Positive Fixed 1 Output 500mA 8-MSOP
        &\href{https://www.digikey.com.au/products/en?keywords=ADP3335ARMZ-5-REELCT-ND}{ADP3335ARMZ-5}
        &Analog Devices Inc.
        &United States\\[4ex]
        IC - 3
        &USB Bridge, USB to UART USB 2.0 UART Interface 28-SSOP
        &\href{https://ftdichip.com/products/ft232rl/}{FT232RL}
        &FTDI, Future Technology Devices International Ltd
        &Scotland\\[4ex]
        LED - 1
        &Ultraviolet (UV) Emitter 405nm 3.2V 30mA 2.5mW/sr @ 20mA 120\textsuperscript{$\circ$} 2-PLCC
        &\href{https://www.digikey.com.au/products/en?keywords=VLMU3100-GS08CT-ND}{VLMU3100-GS08}
        &Vishay Semiconductor Opto Division
        &United States\\[4ex]
        LED - 2
        &Ultraviolet (UV) Emitter 430nm 3.4V 700mA - 130\textsuperscript{$\circ$} 1414 (3535 Metric)
        &\href{https://www.digikey.com/en/products/detail/liteon/LTPL-C034UVH430/7322495}{LTPL-C034UVH430}
        &Lite-On Inc.
        &Taiwan\\[4ex]
        LED - 3
        &Green 505nm LED Indication - Discrete 3.3V 1204 (3210 Metric)
        &\href{https://www.digikey.com.au/products/en?keywords=492-1232-1-ND}{SM1204PGC}
        &Bivar Inc.
        &United States\\[4ex]
        LED - 4
        &Orange 606nm LED Indication - Discrete 2.15V 4-PLCC
        &\href{https://www.digikey.com.au/products/en?keywords=5973209202F-ND}{5973209202F}
        &Dialight
        &United Kingdom\\[4ex]
        LED - 5
        &Red 647nm LED Indication - Discrete 1.7V 0603 (1608 Metric)
        &\href{https://www.digikey.com/en/products/detail/dialight/5975112402F/2432586}{5975112402F}
        &Dialight
        &United Kingdom\\[4ex]
        LED - 6
        &Infrared (IR) Emitter 740nm 1.9V 700mA 250mW/sr @ 350mA 120\textsuperscript{$\circ$} 2-SMD, No Lead Exposed Pad
        &\href{https://www.digikey.com.au/products/en?keywords=1516-1213-1-ND}{QBHP684-IR4BU}
        &QT Brightek (QTB)
        &United States\\[6ex]
        LED - 7
        &Infrared (IR) Emitter 850nm 1.65V 100mA 50mW/sr @ 100mA 20\textsuperscript{$\circ$} 2-SMD, Gull Wing
        &\href{https://www.digikey.com/en/products/detail/vishay-semiconductor-opto-division/VSMY2850G/2615290}{VSMY2850G}
        &Vishay Semiconductor Opto Division
        &United States\\[4ex]
        LED - 8
        &Infrared (IR) Emitter 890nm 1.4V 100mA 6mW/sr @ 100mA 120\textsuperscript{$\circ$} 2-PLCC
        &\href{https://www.digikey.com.au/products/en?keywords=751-1253-1-ND}{VSMF3710-GS08}
        &Vishay Semiconductor Opto Division
        &United States\\[4ex]
        LED - 9
        &Infrared (IR) Emitter 950nm 1.3V 100mA 1.6mW/sr @ 100mA 120\textsuperscript{$\circ$} 2-PLCC
        &\href{https://www.digikey.com.au/products/en?keywords=751-1258-1-ND}{VSMS3700-GS08}
        &Vishay Semiconductor Opto Division
        &United States\\[4ex]
        Microcontroller
        &STM32F051R8T6 Discovery series ARM® Cortex®-M0 MCU 32-Bit Embedded Evaluation Board
        &\href{https://www.st.com/en/evaluation-tools/stm32f0discovery.html}{STM32F0DISCOVERY}
        &STMicroelectronics
        &Switzerland\\[6ex]
        Camera
        &Blackfly S Mono 1.3 MP USB3 Vision (ON Semi PYTHON 1300)
        &\href{https://www.flir.eu/products/blackfly-s-usb3/?model=BFS-U3-13Y3M-C}{BFS-U3-13Y3M}
        &Teledyne FLIR LLC
        &United States\\[4ex]
        Computer
        &
Intel NUC Kit with Intel Celeron Processors
        &\href{https://ark.intel.com/content/www/us/en/ark/products/85254/intel-nuc-kit-nuc5cpyh.html}{NUC Kit NUC5CPYH}
        &Intel Corporation
        &United States\\[2pt]
        \hline
        \hline
        \label{table: component details}
        \end{tabular}
    \end{table}
\end{landscape}

\noindent specimen. The LED driver ICs (MAX16839ASA+) of the illumination panel was powered by an in-house developed AC regulated 12\,V DC power supply and the LEDs were controlled using a separate switching circuit.

\begin{figure*}[t]
\centering
\includegraphics[width=\textwidth]{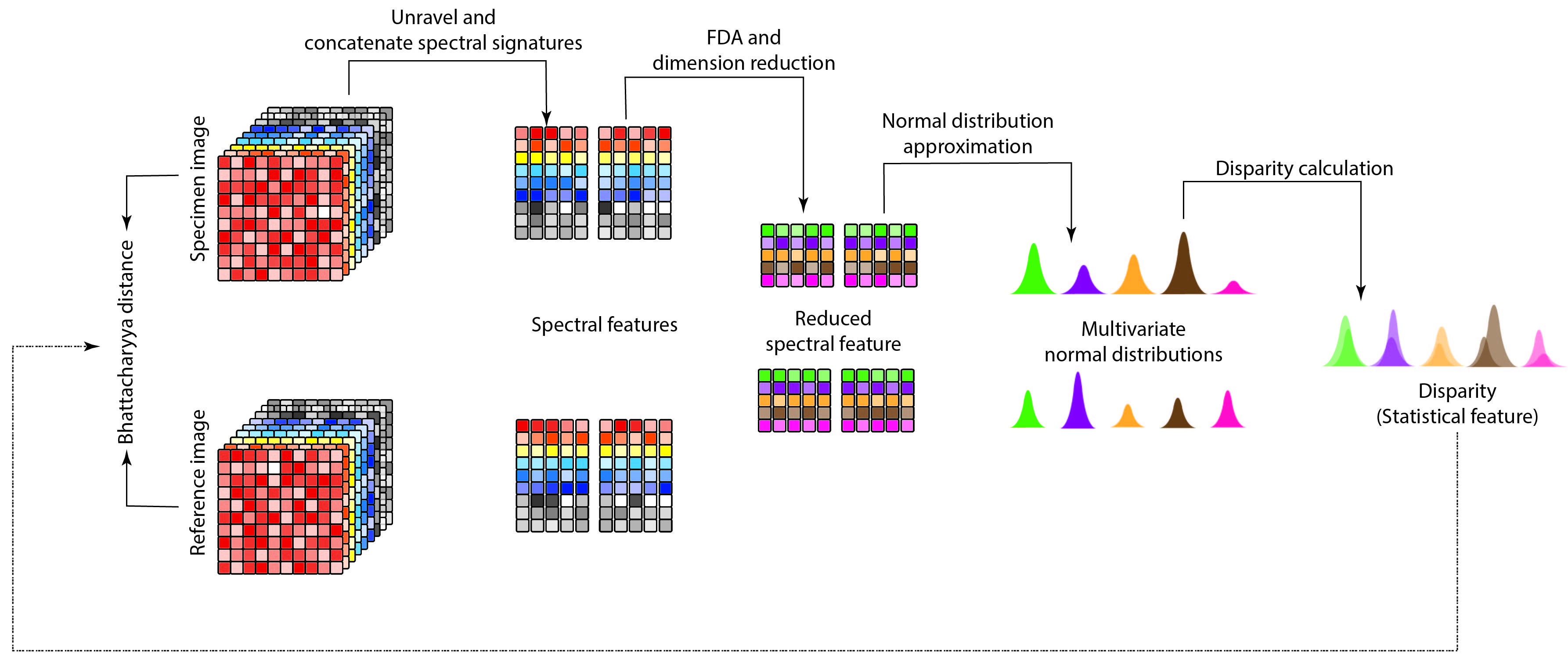}
\caption{Conversion sequence from spectral features to the statistical feature}
\label{figure: conversion sequence}
\end{figure*}

The camera used for the study is a 10-bit CMOS monochrome camera (FLIR Blackfly S Mono, 1.3\,MP, USB3 Vision camera, Resolution – $1280\times1024$) with a similar bandwidth as the illumination panel. A dark chamber was utilized to mount the camera and illumination panel, as well as to hold the liquid sample and to preclude ambient light interference on the sample and the camera.
The dark chamber was fabricated using plywood with dimensions $40\times30\times35\,\text{cm}\textsuperscript{3}$ to affix the LED panel and place the specimen. Also, the design for the dark chamber is given in Fig. \ref{figure: msi design}. 
The camera was mounted at the apex of the chamber to acquire the transmittance spectrum of the sample, and the cost of construction for the proposed imaging setup was USD 650, which is less than that of a low-cost HIS.
A mini PC (Intel\textsuperscript{\textregistered} NUC) was used to store images and to send commands to both the Discovery\textsuperscript{\texttrademark} board (STM32F0DISCOVERY) and the camera. The liquid sample was contained in an opaque cylindrical container with a 10\,cm diameter made of PVC and a 2\,mm plain glass was affixed to the base of the hollow pipe. The monochrome camera and LED switching circuit were synchronized using a Windows batch script before capturing the images.

\subsection{Preparation of samples}
\label{section: preparation of samples}

Authentic, freshly expelled coconut oil was obtained from a reputed large-scale coconut oil producer and exporter. Potatoes were obtained from the local market, washed thoroughly with water to remove soil, peeled off, and sliced into pieces with a commercial slicer to obtain a uniform thickness of 0.5\,cm, and then the slices were cut using a circular cutter to obtain a uniform diameter of 4.5\,cm. The potato slices were blanched at an average temperature of 80\,\textsuperscript{$\circ$}C for 1\,min, blotted with a paper towel (Flora 2-ply kitchen paper towel), packed in zip-lock freezer bags (thickness: 2\,mil, water vapour permeability: 177\,g/\,m\textsuperscript{2}/\,24\,h) and stored at -80\,\textsuperscript{$\circ$}C in food storage refrigerators until further analysis. All the potato samples were processed and stored the day before the initial day of the experiment and used within five days. A batch of potato chips was defrosted for each day of the experiment, and the drip was blotted out before the experiment. On the first day, 1\,L of coconut oil was heated to 150\,\textsuperscript{$\circ$}C and the temperature was maintained in the range of 150\,{$\pm$}\,5\textsuperscript{$\circ$}C for 10\,min by adjusting the flame of the burner. Potato slices (100\,g) were fried in open pans for 3\,min at 150\,{$\pm$}\,5\textsuperscript{$\circ$}C. At the time of the introduction of fresh potato slices into the fryer, it was made sure not to allow the temperature to drop below 140\,\textsuperscript{$\circ$}C. The fried potato slices were removed from the fryer and left to drain for 5\,min under ambient temperature conditions. After allowing oil samples to attain the ambient temperature, an image of repeatedly heated and reused coconut oil was acquired using the multispectral image (MSI) acquisition system. The oil was stored in edible oil cans at ambient temperatures for use on the following day, and the complete process was repeated for five consecutive days. Each day, the number of times an oil sample has been heated was incremented by one, which created a separate reheat cycle count class. The day-to-day reheating done at the consumer level was emulated by the repetitive heating cycles in sample preparation.

\subsection{Image pre-processing}

It is necessary to treat the images to mitigate the noise effects and artefacts superimposed on the captures \cite{qin2013hyperspectral} due to various noise. Hence, the primary correction on the images was to remove any biases added by the camera's sensor. The images were corrected using a dark current reduction step preparatory to further image improvements. In the dark current reduction step, first, an image is captured with zero-illumination in a dark environment (usually known as dark frames / dark current images), and then this dark frame is subtracted from the corresponding actual raw MSIs \cite{porter2008dark}. The dark frames were captured at the beginning of each MSI acquisition process in the implementation. After that, the dark current subtraction process was applied to subsequent MSIs utilizing the \ref{equation: dark current removal},

\begin{equation}
\label{equation: dark current removal}
P[\lambda]=S[\lambda]-D
\end{equation}
where, $P[\lambda]$ is the dark current removed image at wavelength $\lambda$, $S[\lambda]$ is the raw image at wavelength $\lambda$ and $D$ is the dark current image.
However, it is impossible to perform the spectral image normalization using white-references, which is the standard practice for reflectance-based MSIs \cite{qin2013hyperspectral} for transmittance-based due to physical constraints. Even though the white-reference normalization in the reflectance configuration can increase the resolution of the pixel values through the scaling operation, it is ineffective in the transmittance configuration as more light is passed through the specimen, thus yielding normalized values closer to unity.

Next, the nonlinear median filtering process was carried out on the dark current subtracted images to remove the aforementioned random noise from the images. Here the median filter was utilized to remove some inherent noise by removing the isolated pixels while preserving the spatial resolution \cite{acharya2005image}. The input pixel value is replaced by the moving average filter output given by,

\begin{equation}
\label{equation: noise filtering}
P^{*} [i,j] = \frac{1}{N}\sum_{k = i-w}^{i+w} \sum_{l = j-w}^{j+w} P [k,l]
\end{equation}
where, $P^{*} [i,j]$ is the replaced value of the pixel $(i,j)$, $P[k,l]$ is the pixel value of the dark current subtracted image at $(k,l)$ location, $w$ is a suitable window size, and $N$ is the number of pixels in the window. A window size of $30 \times 30 $ is chosen for the implementation of \ref{equation: noise filtering} in the application.

\subsection{Algorithms on reheated oil analysis}
\label{section: algorithms on reheated oil analysis}

\subsubsection{Dataset}
\label{section: dataset}

For the experiment, nine independent trials were conducted where the oil sample was reheated six times during each trial. In each trial, an MSI was acquired at the end of reheating for each reheating instance. After pre-processing the image, a $30\times30$ window was cropped, which results in 900 spectral signatures for the respective reheat cycle count class under the respective trial, and when accounted for the nine trials, 8100 spectral signatures were recorded for each reheat cycle count class. Since there were six classes, a dataset with a size of 48600 spectral signatures were recorded and made available for public access \cite{weerasooriya2020mendeley}. This dataset was used for both studies, nonetheless, with the reformation of the dataset as appropriately.

For the first study in \ref{section: reheat cycle estimation}, labelled sets are required to train the classifier with the reheat cycle count as the class label. According to related work on using multispectral images to detect oil-containing salad dressing \cite{raju2020detection}, a sample size of 44.6 is sufficient for each class to produce a statistically significant result according to \cite{cohen1992statistical}. Hence, sixty labelled sets were prepared from the 8100 spectral signatures recorded for each reheat cycle count class. Consequently, each labelled set included 135 spectral signatures once the 8100 signatures of each reheat cycle count class were divided amongst the sixty sets, and the reformed dataset consisted of a total of 360 labelled sets. Next, a training-testing split of 80:20 was used on the sixty labelled sets from each reheat cycle count class.

For the second study in \ref{section: discrimination of appreciable alterations}, the proposed method identifies significant property changes in the same oil sample as the reheat cycle count is increased through unsupervised classification of the multispectral images. Since the image data from different trails should not be mixed and the learning technique is unsupervised, the labelled sets used for the first study can not be used here. Instead, the original dataset was separated according to their respective trial to create nine subsets of data, and each subset included the spectral signatures for all six reheat cycle count classes of that trial. Accordingly, each subset, therefore, constituted a total of 5400 spectral signatures.

\subsubsection{Reheat cycle count estimation}
\label{section: reheat cycle estimation}

The first objective of the study was to estimate the number of times a given coconut oil sample has been reheated as compared to a pure sample. Since the estimated quantity is a discrete variable, the estimation problem can be converted to a classification problem. To construct the classifier, first, the dimension of the training dataset was reduced with FDA. Even though the classifier could be trained to output the estimated reheat cycle count of each pixel --- with the reduced spectral signature at each pixel considered as the input --- and then attempt to interpret the reheat cycle count class of the image in a post-analysis, it is more pertinent to determine the reheat cycle count of the entire oil sample at once with the input image for the classifier. For that, the input to the classifier has to be constructed by concatenating the individual spectral signatures of each sample. However, this circumscribes the selection of an appropriate sample size of choice. To that end, the classifier was developed to use the statistics of the sample rather than raw signatures, and the use of statistics relieves the constraints on the sample size and the requirement of a post-analysis of a pixel-level classification.

Here, a separate pure oil sample was used as the reference sample to compute Bhattacharyya distance, and each reheated oil sample was a target sample. Bhattacharyya distance ($D_B$) is a measure of the similarity between two distributions and can be calculated according to \ref{equation: bhattacharyya distance} for multivariate normal distributions,

\begin{align}
    D_{B_1} = & ~\frac{1}{8}(\mathbf{\mu_t - \mu_r})^{\top}\left(\frac{\mathbf{\Sigma_t + \Sigma_r}}{2}\right)^{-1}(\mathbf{\mu_t - \mu_r})\\
    D_{B_2} = & ~\frac{1}{2}\log\left(\frac{\det\left(\frac{\mathbf{\Sigma_t + \Sigma_r}}{2}\right)}{\sqrt{\det\mathbf{\Sigma_t}\det\mathbf{\Sigma_r}}}\right)\\
\label{equation: bhattacharyya distance}
    D_B = & ~ D_{B_1} + D_{B_2}
\end{align}
with $\mathbf{\mu_t}$, $\mathbf{\Sigma_t}$ as the mean signature and the covariance matrix of the target oil sample, and $\mathbf{\mu_r}$, $\mathbf{\Sigma_r}$ as the mean signature and the covariance matrix of the reference oil sample. Then $D_{B_1}$ and $D_{B_2}$ represent the Mahalanobis distance and variance disparity between the two distributions.
With the calculation of Bhattacharyya distance ($D_B$), the spectral features of the oil samples will be converted to a statistical feature and this uni-dimensional feature is used for the classification of reheat count classes. The conversion sequence from spectral features to the statistical feature is summarized in Fig. \ref{figure: conversion sequence}.

For the classification of the sample, a Support Vector Machine (SVM) \cite{scikit-learn, cortes1995support} classifier was used to find the optimal decision thresholds with Bhattacharyya distance. The resulting linear decision boundaries from the use of a single input espoused the use of an SVM classifier, as the SVM classifier is superior at finding the best hyperplanes which maximally separate different classes (see Appendix \ref{appendix: A}). An SVM classifier with the radial basis function as the kernel was constructed using five-fold cross-validation. The parameter $\gamma$ and the cost were chosen through a parametric sweep (see Appendix \ref{appendix: B}) for the optimal combination of the two parameters. The SVM model was evaluated for all data points using the accuracy of the classification, and since the dataset used for training was balanced, the accuracy was defined as in \ref{equation: svm classification accuracy} as below,

\begin{equation}
\label{equation: svm classification accuracy}
Accuracy = \frac{TP + TN}{TP + TN + FP + FN}
\end{equation}
with the terms true-positive (TP), true-negative (TN), false-positive (FP), and false-negative (FN).

\begin{figure*}[t!]
    \centering
%\begin{sidewaysfigure}
    \begin{subfigure}[t]{0.75\textwidth}
        \centering
        \includegraphics[width=\textwidth, angle=0, ]{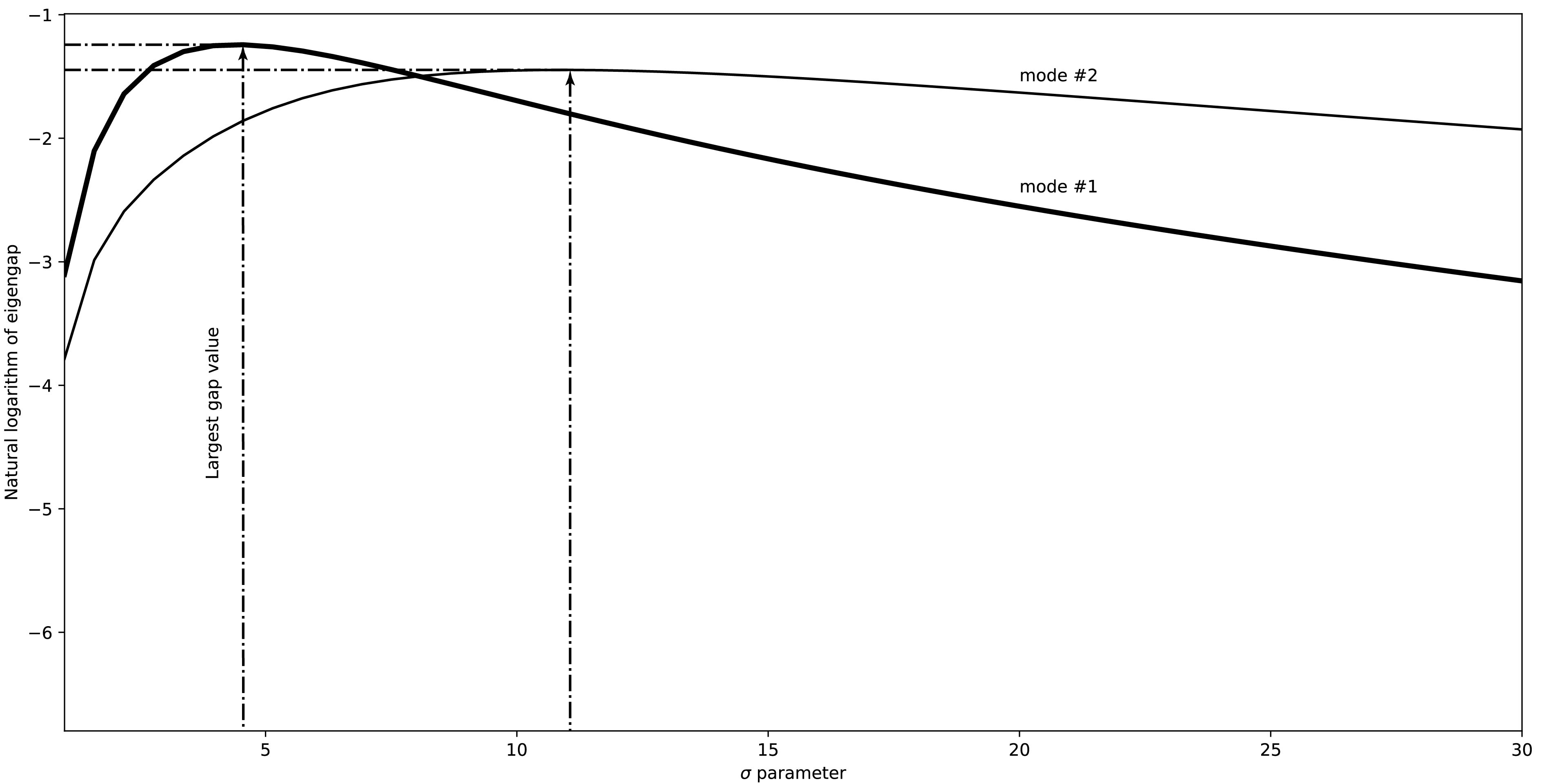}
        \captionsetup{justification=centering}
        \caption{}
        \label{figure: lgv criterion}
    \end{subfigure}
    \medskip
    \begin{subfigure}[t]{.75\textwidth}
        \centering
        \includegraphics[width=\textwidth, angle=0,]{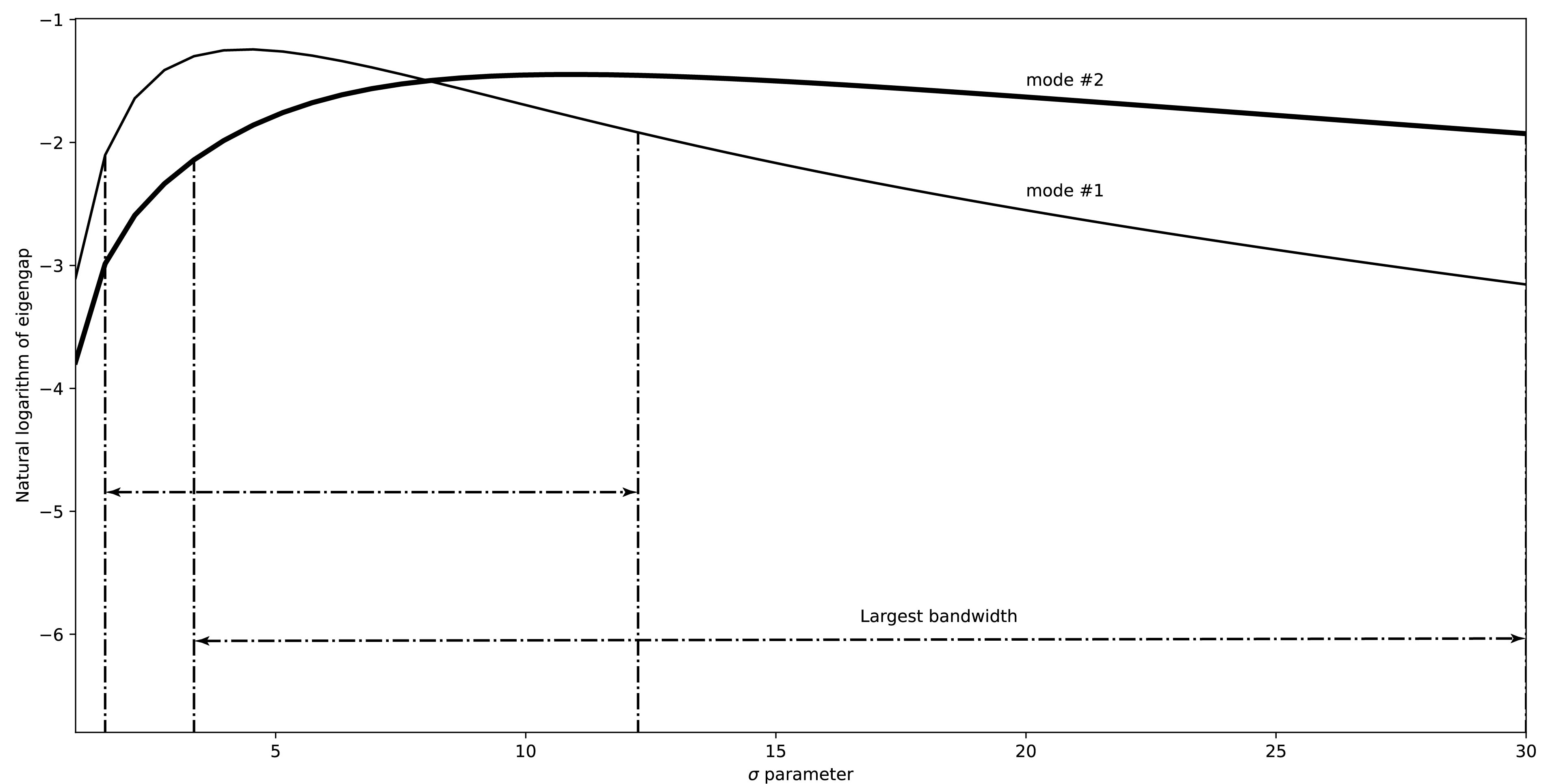}
        \captionsetup{justification=centering}
        \caption{}
        \label{figure: lbw criterion}
    \end{subfigure}
    
%\end{sidewaysfigure}
    \caption{Prominent mode selection under (a) largest gap value and (b) largest bandwidth algorithms (sigma-sweep curve of the prominent mode is thickened)}
    \label{figure: spectral clustering algorithms}
\end{figure*}

\subsubsection{Discrimination of appreciable alterations}
\label{section: discrimination of appreciable alterations}

The number of reheating cycles of an oil sample is a discrete realization of an underlying continuous process, vis-\`a-vis, changes in chemical properties. Even though the control parameters (temperature and duration) were maintained uniform throughout different trials, frying of different potato batches could change the chemical properties of the oil gradually or abruptly at consecutive reheats. Therefore, identifying any sudden change in oil properties is an indicator of significant damage to the oil with a step increment in the reheat cycle count. However, the development of the discriminator falls within the realms of unsupervised learning because the developed framework has to be applicable in the case of other fryable foods. Since the framework's objective is to realize reheat cycle count classes where a noticeable change has occurred in the oil, the problem could be transformed into an unsupervised clustering problem, wherein only gradual changes have had happened with the number of reheat cycle counts in each cluster. However, to group these reheat cycle count classes, it is imperative to determine the optimal number of clusters considering the global structure of the spectral signatures prior to clustering. To that end, spectral clustering (SC) \cite{ng2001spectral} is an apposite clustering algorithm that facilitates the consideration of the global structure of the dataset and the search for the optimal number of clusters in the dataset. SC uses the spectral connectivity of the dataset to group data; hence the connectivity amongst data points has to be improved for optimal grouping. Sigma-sweep \cite{rupasinghe2016modes} was performed to augment the spectral connectivity,  according to \ref{equation: laplacian matrix} as below, 

\begin{align}
\label{equation: laplacian matrix}
    \mathbf{W}[i,j] &=
    \begin{cases}
      \exp({-\frac{||x_{i}-x_{j}||^2}{2 \sigma^{2}}}) &; i \neq j \\
      0 \hphantom{0000000000000} &; \text{otherwise}
    \end{cases}\\
    \mathbf{D}[i,k] &=
    \begin{cases}
      \sum_{j} \mathbf{W}[i,j] &; i = k \\
      0 \hphantom{0000000000000} &; \text{otherwise} 
    \end{cases}\\
    \mathbf{L} &= \mathbf{I}-\mathbf{D}^{-\nicefrac{1}{2}}\,\mathbf{W}\, \mathbf{D}^{-\nicefrac{1}{2}}\text{,}
\end{align}

where, $\mathbf{W}$ is the affinity matrix, $\mathbf{D}$ is the degree matrix, $\mathbf{L}$ is the Graph Laplacian and $\mathbf{I}$ is the identity matrix. A Gaussian kernel was used to compute the affinity matrix, with $x_i$ and $x_j$ denoting the $i\textsuperscript{th}$ and $j\textsuperscript{th}$ signatures of the set of spectral signatures for the oil sample and $\sigma$ as the radius of neighbourhood. Then, $\mathbf{Y}[m,n]$ denotes the matrix element at the $m$\textsuperscript{th} row and $n$\textsuperscript{th} column of the matrix $\mathbf{Y}$.

In the sigma-sweep, the tuning parameter ($\sigma$) was increased from $1.0$ since the requirement was to group different reheat cycle count classes rather than detecting sub-clusters within a class. This grouping method will lump classes with marginal changes in spectral properties together while separating reheat cycle count class clusters with significant changes. Accordingly, the outcome of the grouping is the formation of umbrella clusters of classes with incremental changes in spectral properties — which reflects the chemical characteristics of the oil — due to reheating. Hence, the inter-class boundaries demarcate significant or perhaps critical changes in composition and are worthy of identification because of the health risks. Then, to determine the optimal number of clusters, the eigengaps were considered between the third and sixth eigengap, where eigengap is the difference between consecutive eigenvalues of $\mathbf{L}$ in the ascending order.

Two selection algorithms were used to select the optimal number of clusters, considering the variation of the sigma-sweep curves. 

\begin{enumerate}
    \item Largest gap value (LGV): the eigengap with the highest gap value was considered as the prominent mode. The connectivity of the spectral signatures is at its summit for the resulting prominent mode.
    \item Largest bandwidth (LBW): the eigengap with the largest $\sigma$ span for which the gap values are higher than half the maximum gap value for that mode was considered as the prominent mode. The connectivity of the spectral signatures is more stable in the prominent mode than the structures resulting from the rest of the modes.
\end{enumerate}

Once the prominent mode is selected, the $\sigma$ value for which the highest gap value was recorded for the prominent mode was selected as the dominant $\sigma$ under each selection algorithm. The dominant $\sigma$ improves the spectral connectivity, and the prominent mode gives the optimal number of clusters. In Fig. \ref{figure: spectral clustering algorithms}, the curve characteristics corresponding to the two selection algorithms are demonstrated. Once these two parameters (prominent mode and dominant $\sigma$) are computed, the reheat cycle count classes are clustered using SC to distinguish the classes that have had a marked change in the spectral properties from one reheat cycle count class to the next.

\renewcommand{\thesection}{\Roman{section}}

\section{Results}
\label{section: results}

\subsection{Multispectral imagery \textit{vs.} RGB photography}
\label{section: multispectral images and rgb photography}

\begin{figure}[t!]
    \centering
    \begin{subfigure}[t]{0.45\textwidth}
        \centering
        \captionsetup{justification=centering}
        \includegraphics[width=1.0\textwidth]{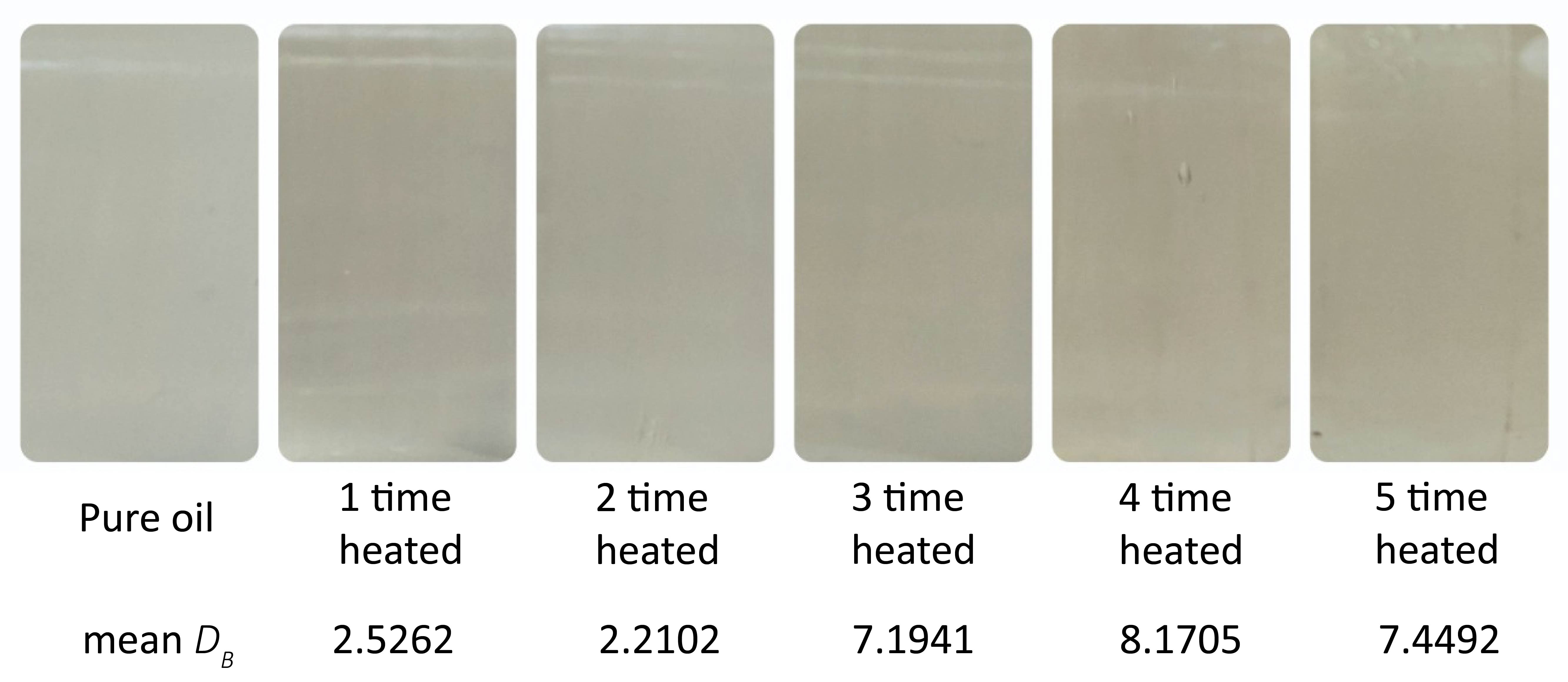}
        \caption{}
        \label{figure: rgb images of oil}
    \end{subfigure}
    %~\hspace{10mm} 
    \begin{subfigure}[t]{0.45\textwidth}
        \centering
        \captionsetup{justification=centering}
        \includegraphics[width=1.0\textwidth]{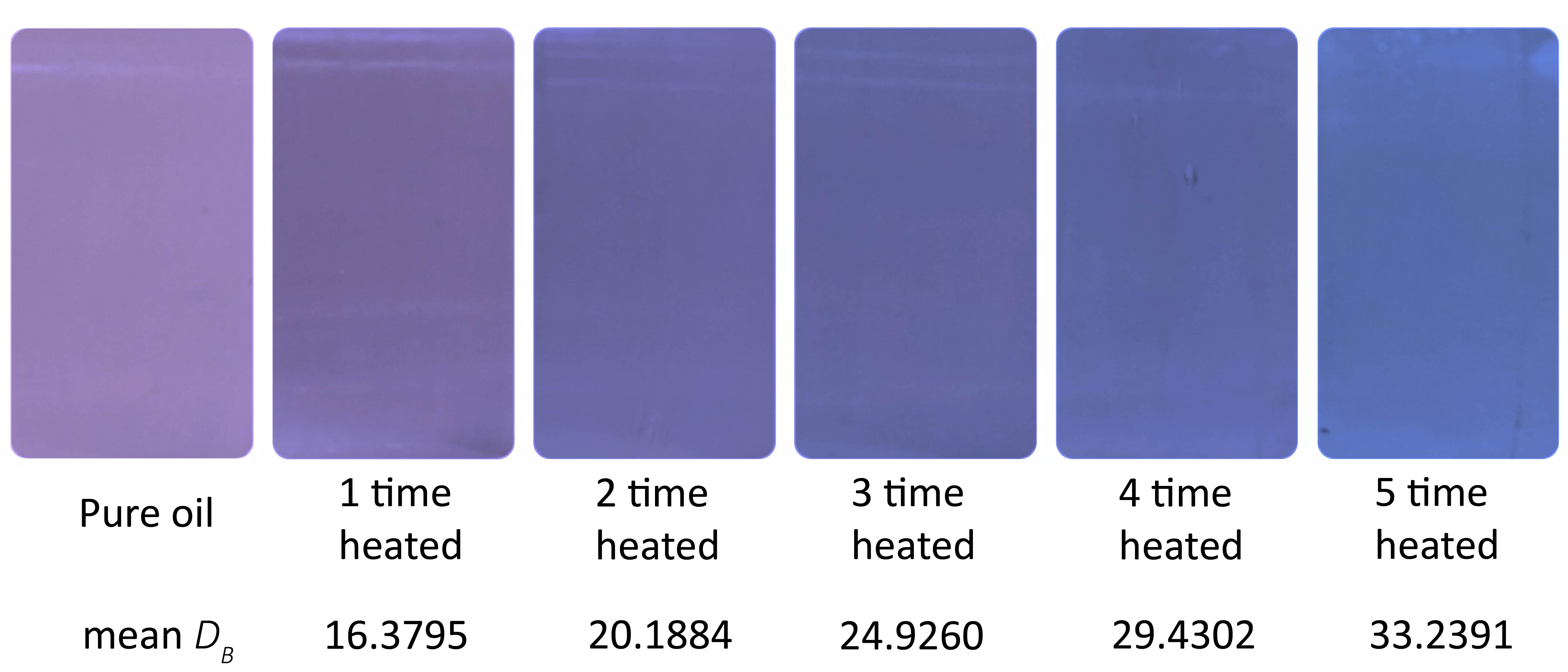}
        \caption{}
        \label{figure: false rgb images of oil}
    \end{subfigure}
    \caption{(a) True RGB images and (b) false RGB representation of MSIs of reheated oils}
    \label{figure: Spectral images of oil}
\end{figure}

The premise for the application of image analysis is that the changes in the oil properties are observable by the spectral properties of the oil sample. For illustration, the true-colour image under RGB photography is given in Fig. \ref{figure: rgb images of oil} of the same oil batch for different reheat cycle count classes. The false RGB representation of the MSIs of the same oil batch is provided in Fig. \ref{figure: false rgb images of oil} to demonstrate the differentiation capability of the MIS, as the raft of dimensions of MSIs are cumbersome to be showcased. Also, the mean Bhattacharyya distance to each reheat cycle count class from the pure oil class is given under each class for both the imaging techniques.

\subsection{Classifier decision boundaries and performance}
\label{section: classifier decision boundaries and performance}

\begin{figure*}[ht!]
    \centering
    \begin{subfigure}[t]{\columnwidth}
        \centering
        \captionsetup{justification=centering}
        \includegraphics[height=0.7\textwidth]{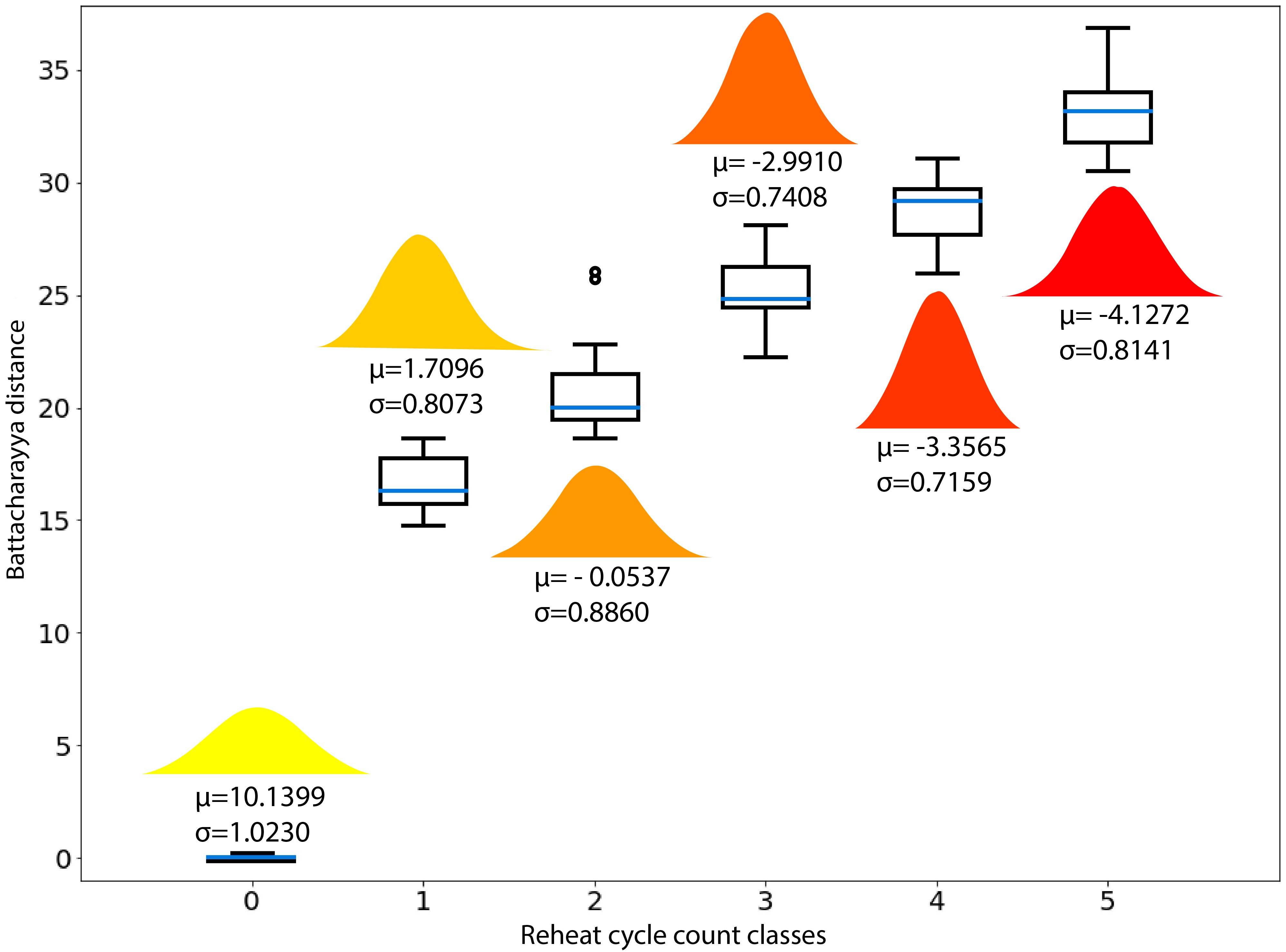}
        \caption{}
        \label{figure: variation of bhattacharyya distances}
    \end{subfigure}
    ~ 
    \begin{subfigure}[t]{\columnwidth}
        \centering
        \captionsetup{justification=centering}
        \includegraphics[height=0.7\textwidth]{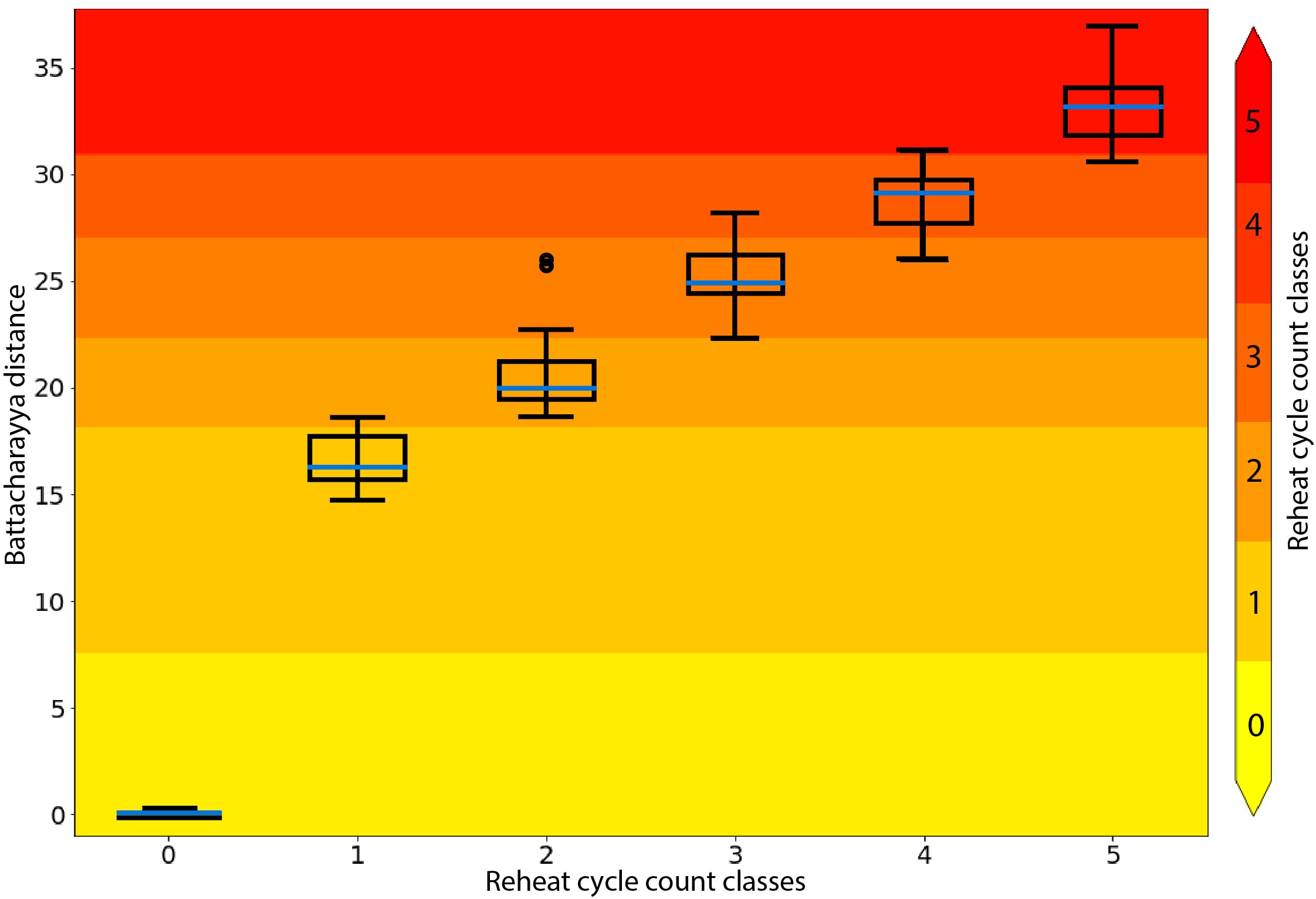}
        \caption{}
        \label{figure: decision boundaries}
    \end{subfigure}
    \caption{
    (a) Variation of Bhattacharyya distances with reheat cycle count classes with the variation of the average Gaussian distribution along the first eigenvector and (b) decision boundaries of the SVM classifier for class estimation
    }
    \label{figure: bhattacharyya distances and decision boundaries}
\end{figure*}

\begin{figure*}[ht!]
    \centering
    \begin{subfigure}[t]{\columnwidth}
        \centering
        \captionsetup{justification=centering}
        \includegraphics[height=1.0\textwidth, width=1.0\textwidth]{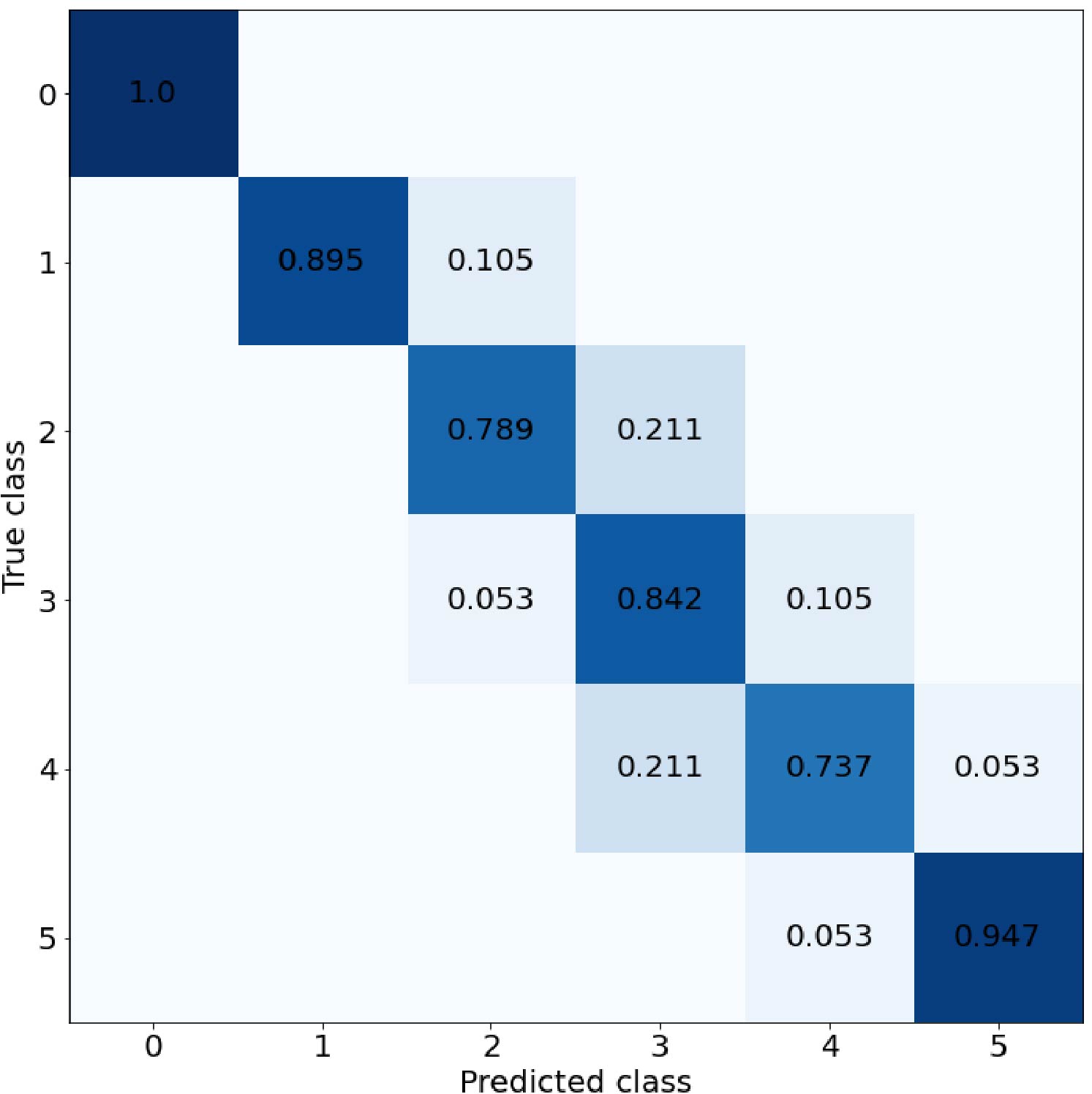}
        \caption{}
        \label{figure: confusion matrix for train data}
    \end{subfigure}
    ~ 
    \begin{subfigure}[t]{\columnwidth}
        \centering
        \captionsetup{justification=centering}
        \includegraphics[height=1.0\textwidth, width=1.0\textwidth]{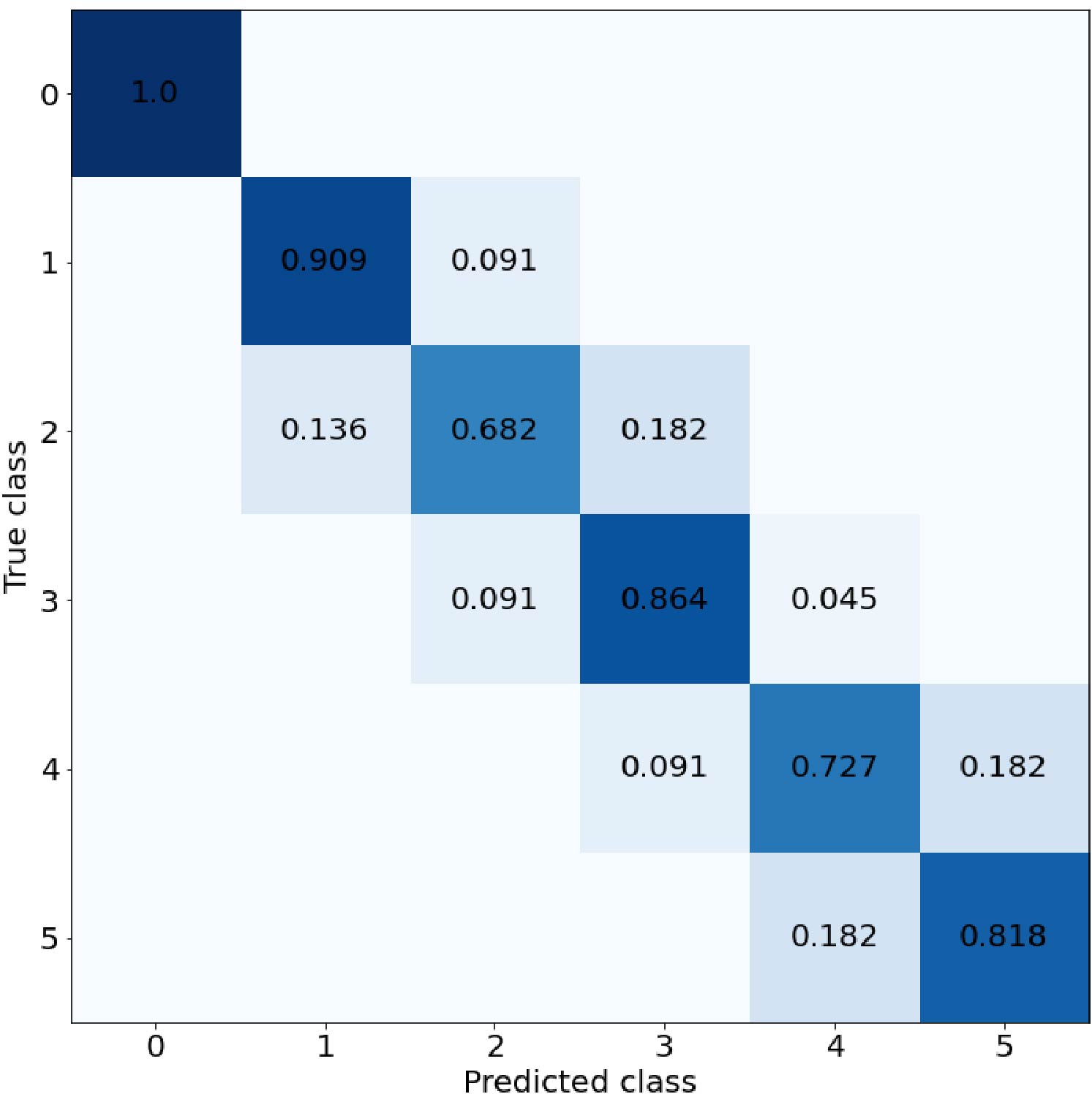}
        \caption{}
        \label{figure: confusion matrix for test data}
    \end{subfigure}
    \caption{Classification accuracy of reheat cycle count classes estimation for (a)train and (b)test data }
    \label{figure: classification accuracies}
\end{figure*}

The SVM classifier was developed with the Bhattacharyya distance between an oil sample and the pure sample (reference sample) as the input variable. The variation of Bhattacharyya distances computed for different reheat cycle count classes are given Fig. \ref{figure: variation of bhattacharyya distances}, and the corresponding decision boundaries of the SVM classifier are depicted in Fig. \ref{figure: decision boundaries}. The classification results for the reheat cycle count class estimation for train and test data are presented in the form of confusion matrices in Fig. \ref{figure: classification accuracies}, respectively. The box plots given in Fig. \ref{figure: decision boundaries} are mostly contained within the decision boundaries from the SVM classifier, which is propitious for higher classification accuracies as corroborated by the confusion matrices in Fig. \ref{figure: classification accuracies}, whose non-zero elements are primarily on either the main or the closest diagonal.

\subsection{Sigma-sweep and reheat cycle count classes clustering}
\label{section: sigma sweep and reheat cycles clustering}

\begin{figure*}[ht!]
    \centering
\begin{subfigure}{\columnwidth}

\begin{subfigure}{\columnwidth}
  \captionsetup{justification=centering}
  \includegraphics[width=\linewidth]{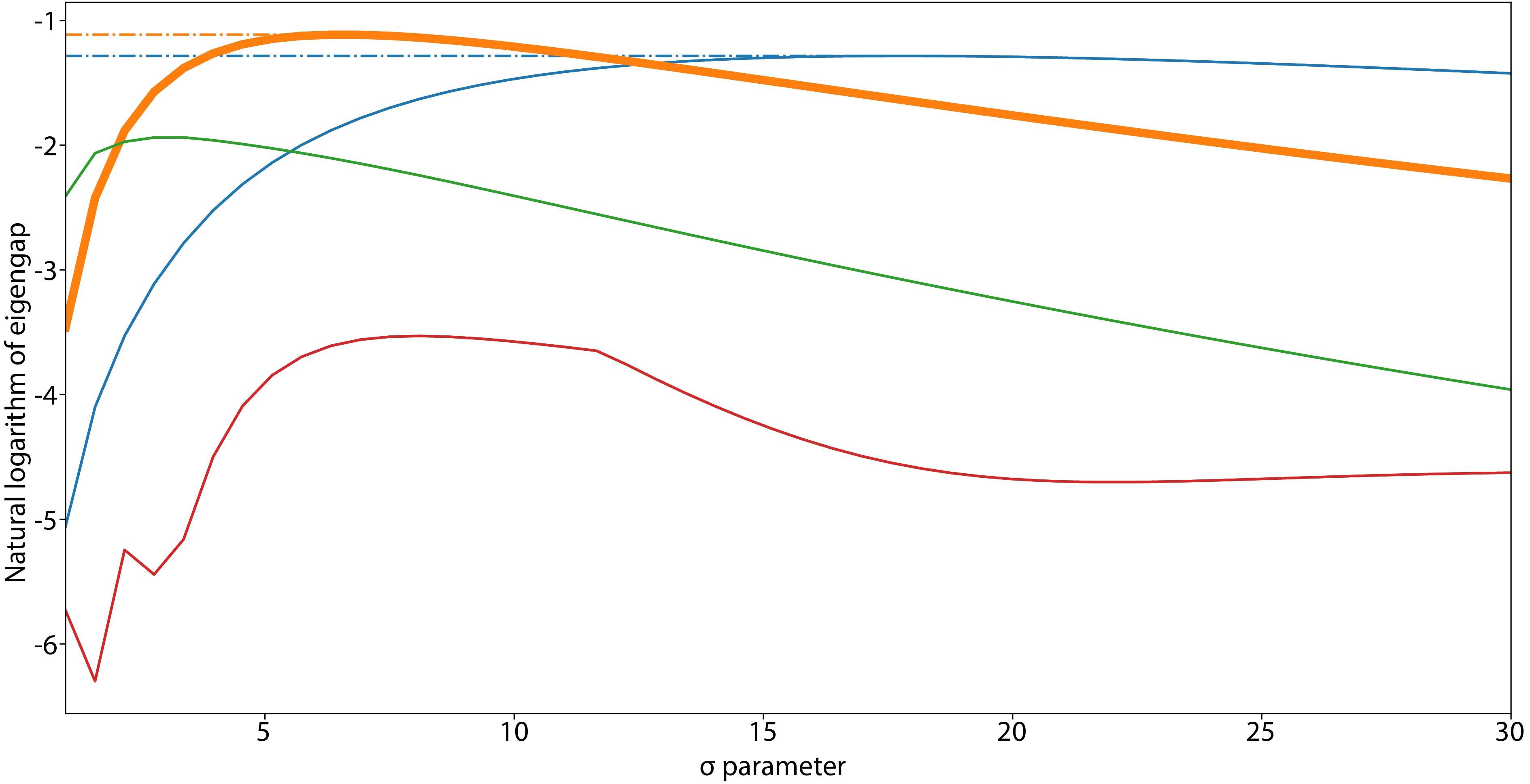}
  \caption{}
  \label{figure: lgv mode selection trial 0}
\end{subfigure}
\medskip
\begin{subfigure}{\columnwidth}
  \captionsetup{justification=centering}
  \includegraphics[width=\linewidth]{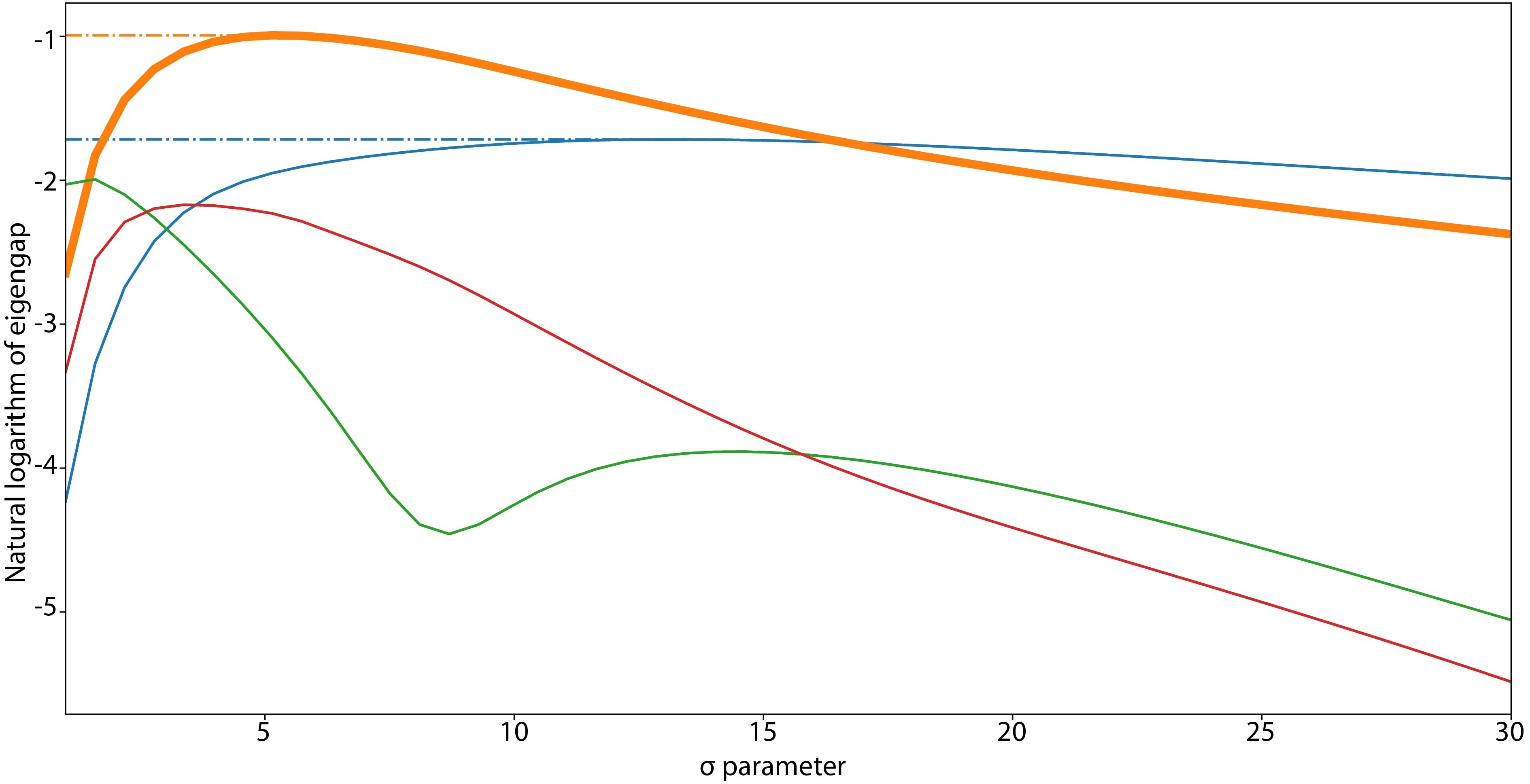}
  \caption{}
  \label{figure: lgv mode selection trial 5}
\end{subfigure}
\medskip
\begin{subfigure}{\columnwidth}
  \captionsetup{justification=centering}
  \includegraphics[width=\linewidth]{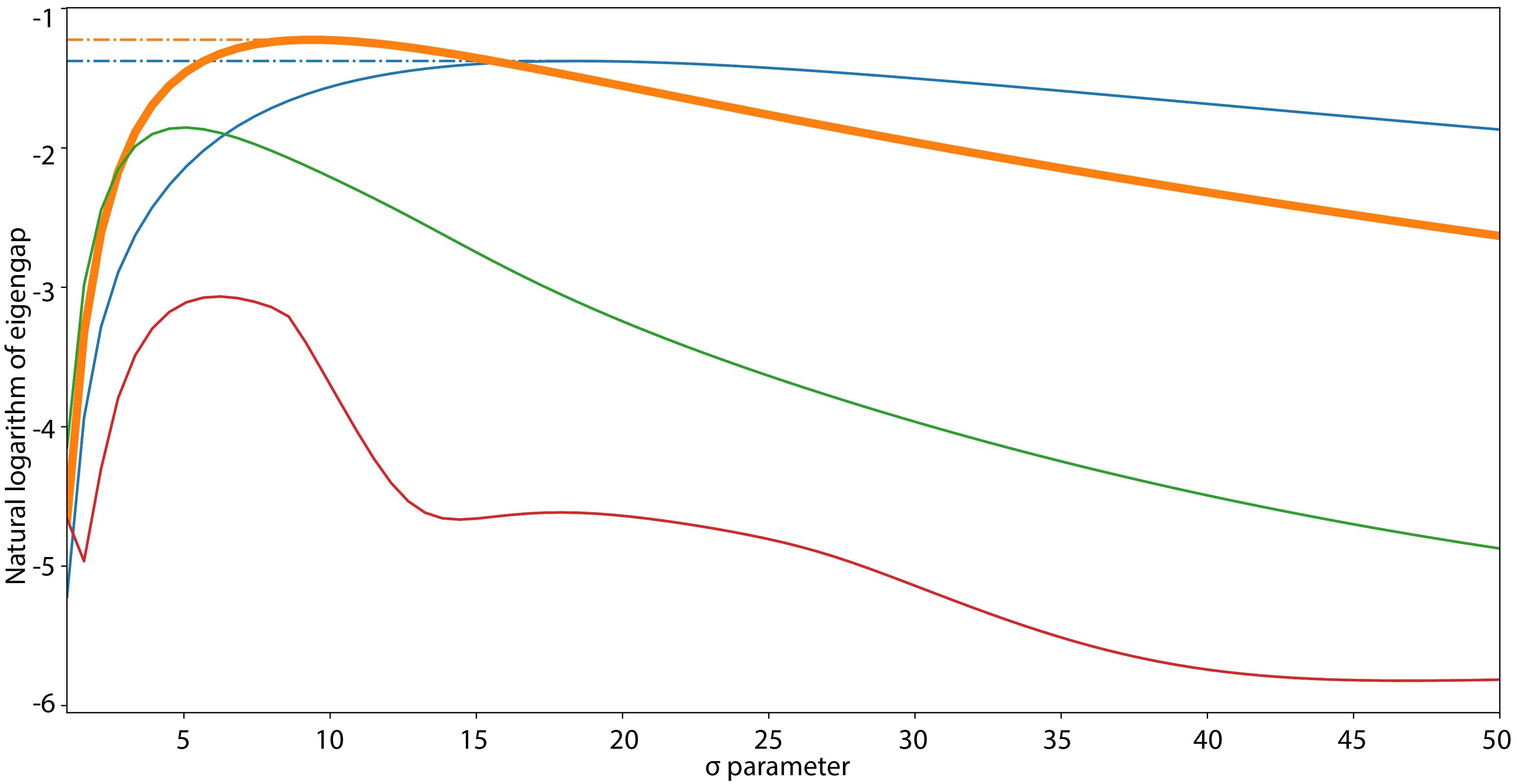}
  \caption{}
  \label{figure: lgv mode selection overall trials}
\end{subfigure}    

\end{subfigure}
~~
\begin{subfigure}{\columnwidth}

\begin{subfigure}{\columnwidth}
  \captionsetup{justification=centering}
  \includegraphics[width=\linewidth]{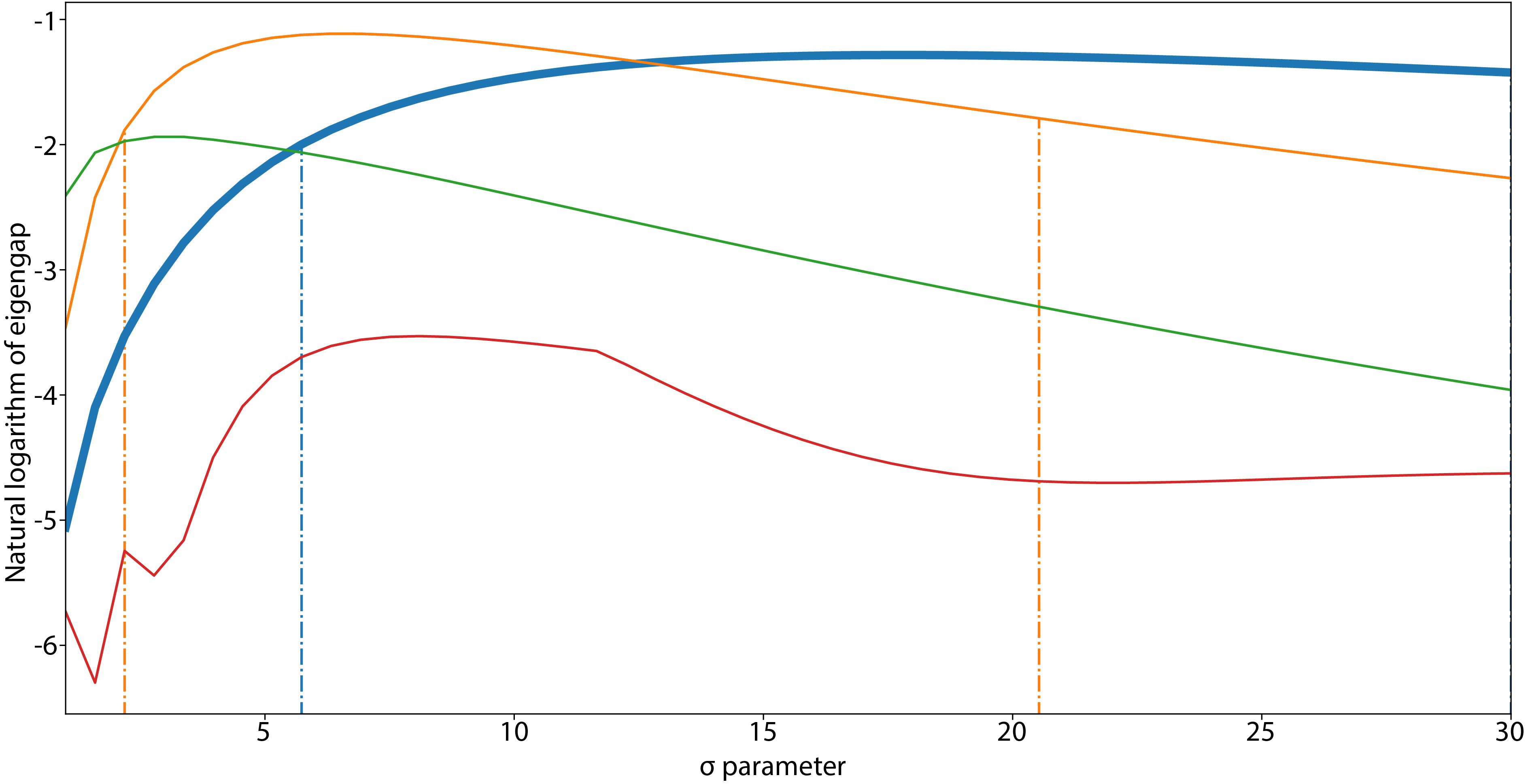}
  \caption{}
  \label{figure: lbw mode selection trial 0}
\end{subfigure}
\medskip
\begin{subfigure}{\columnwidth}
  \captionsetup{justification=centering}
  \includegraphics[width=\linewidth]{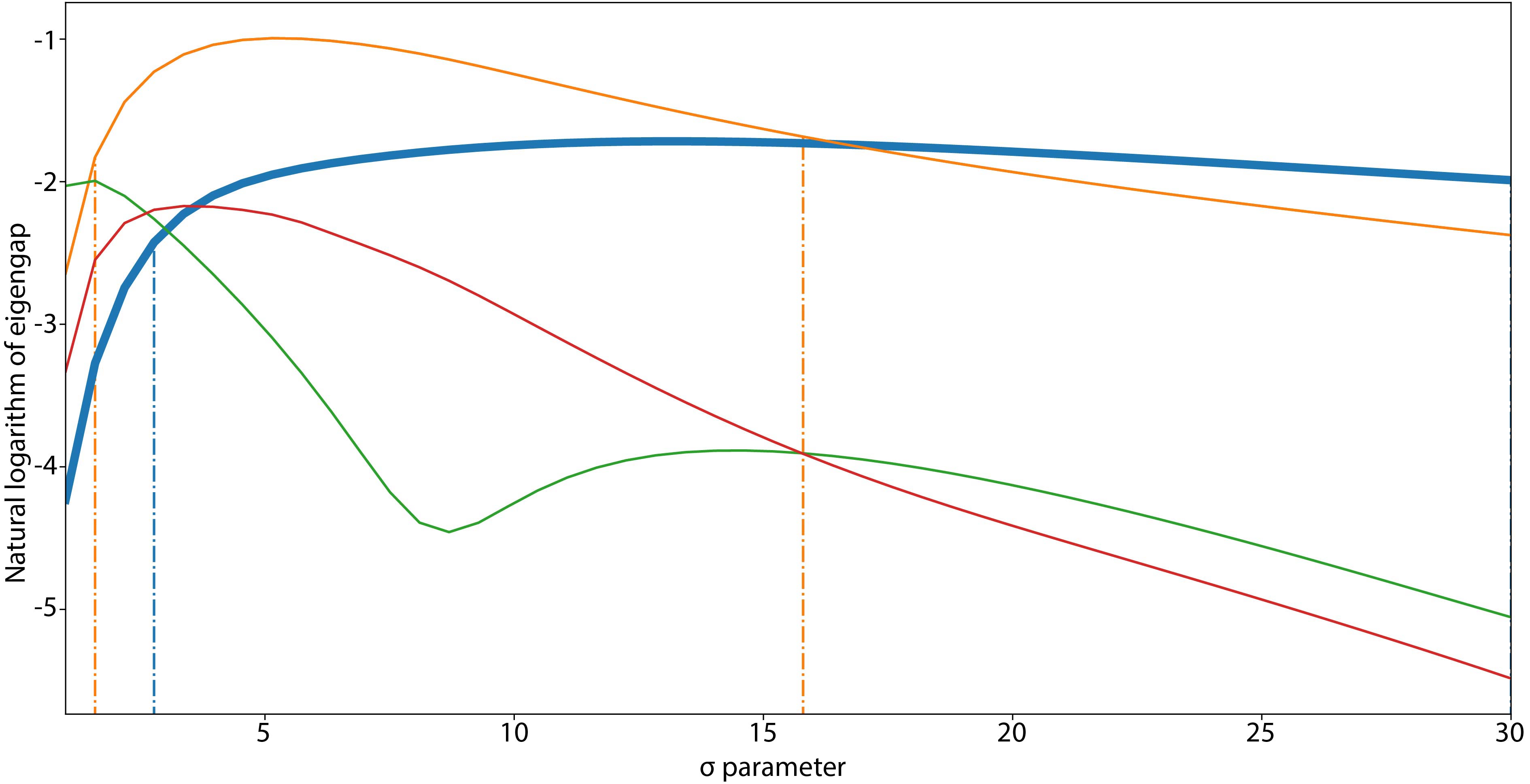}
  \caption{}
  \label{figure: lbw mode selection trial 5}
\end{subfigure}
\medskip
\begin{subfigure}{\columnwidth}
  \captionsetup{justification=centering}
  \includegraphics[width=\linewidth]{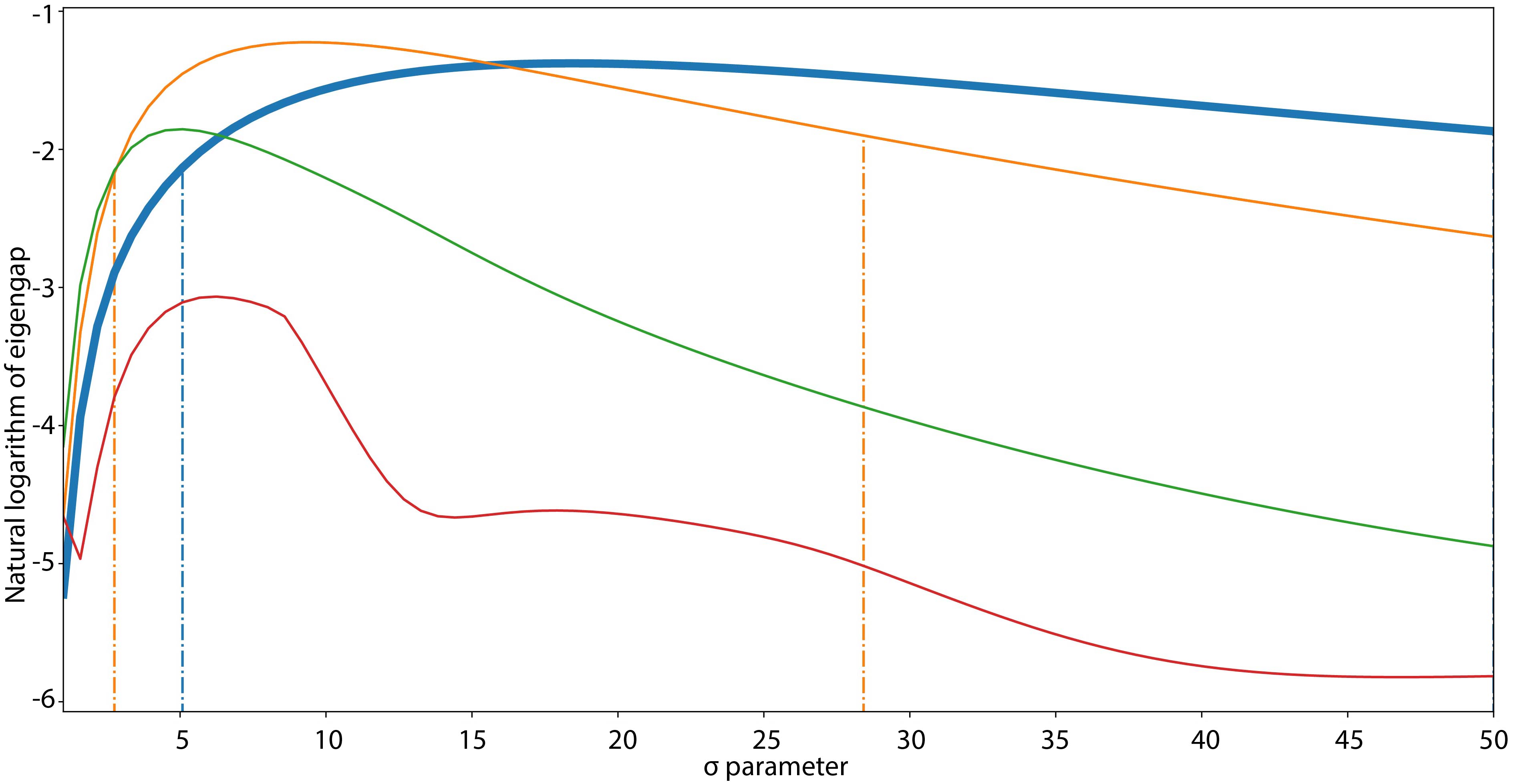}
  \caption{}
  \label{figure: lbw mode selection overall trials}
\end{subfigure}    

\end{subfigure}
\medskip
\begin{subfigure}{0.5\textwidth}
    \includegraphics[width=\textwidth]{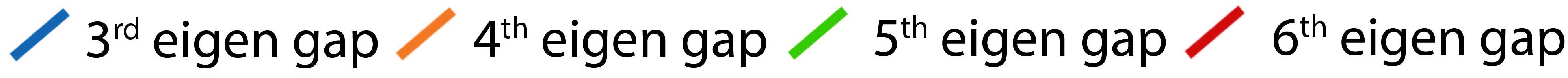}
    \captionsetup{justification=centering}
    \caption*{Legend for sigma-sweep curves}
\end{subfigure}

\caption{
Mode selection under LGV algorithm for (a) trial $0$, (b) trial $5$, and (c) overall trials and under LBW algorithm for (d) trial $0$, (e) trial $5$, and (f) overall trials with sigma-sweep curves (prominent mode curve is thickened)}
\label{figure: mode selection with sigma sweep curves}
\end{figure*}

\begin{figure*}[ht!]
    \centering
\begin{subfigure}{0.33\textwidth}
  \captionsetup{justification=centering}
  \includegraphics[width=\linewidth]{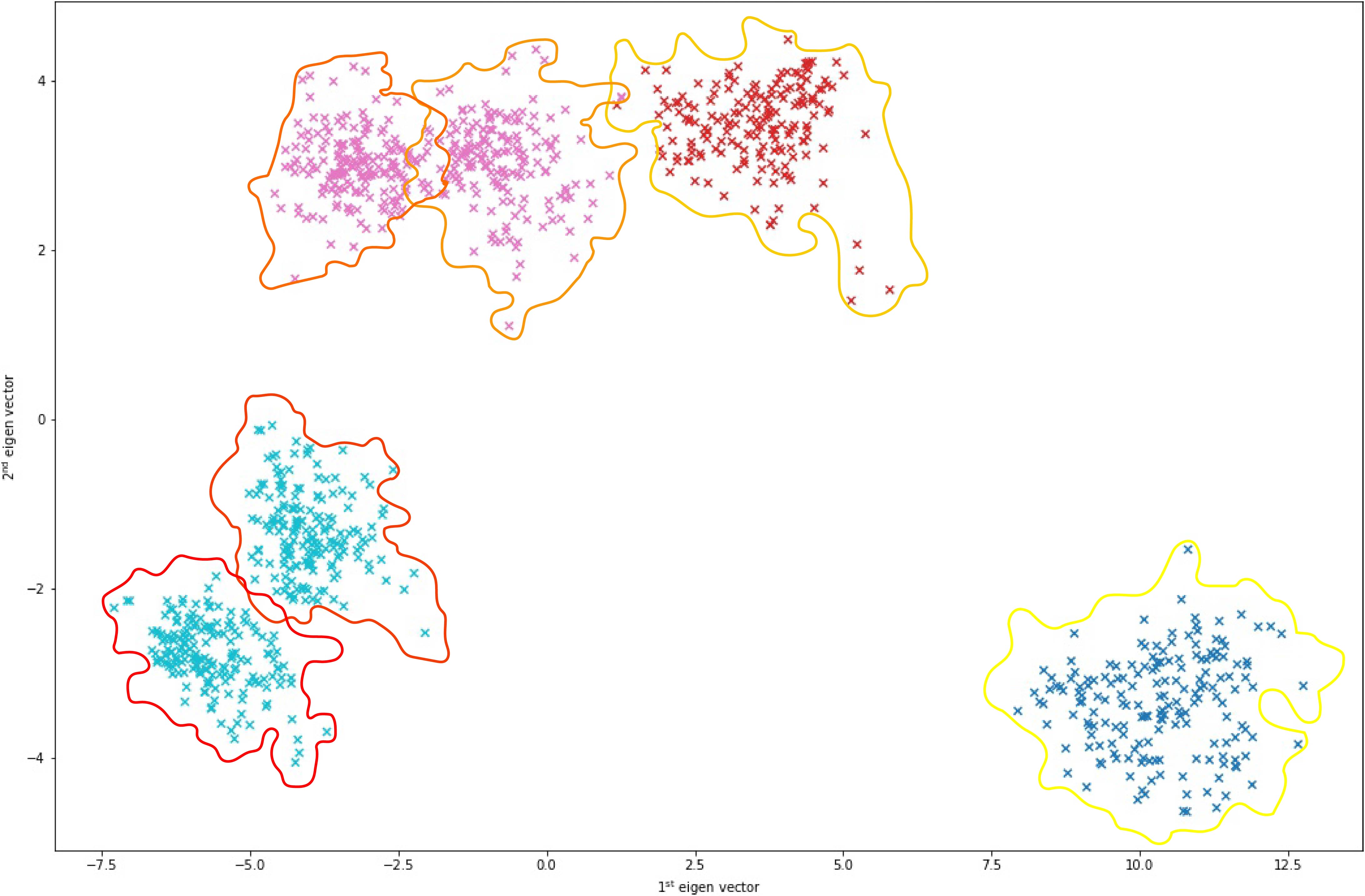}
  \caption{}
  \label{figure: lgv clusters trial 0}
\end{subfigure}
\begin{subfigure}{0.33\textwidth}
  \captionsetup{justification=centering}
  \includegraphics[width=\linewidth]{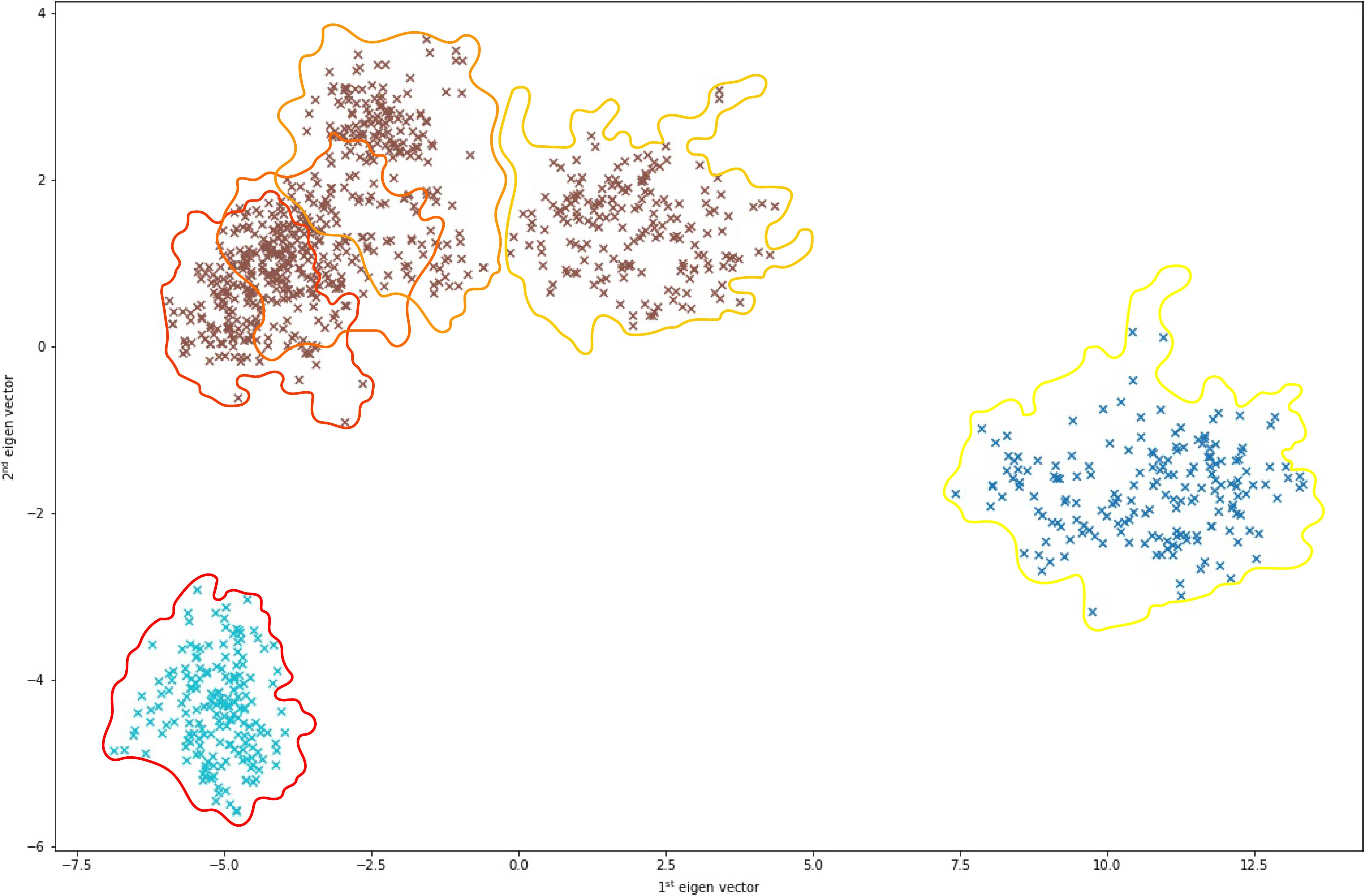}
  \caption{}
  \label{figure: lgv clusters trial 5}
\end{subfigure}
\begin{subfigure}{0.33\textwidth}
  \captionsetup{justification=centering}
  \includegraphics[width=\linewidth]{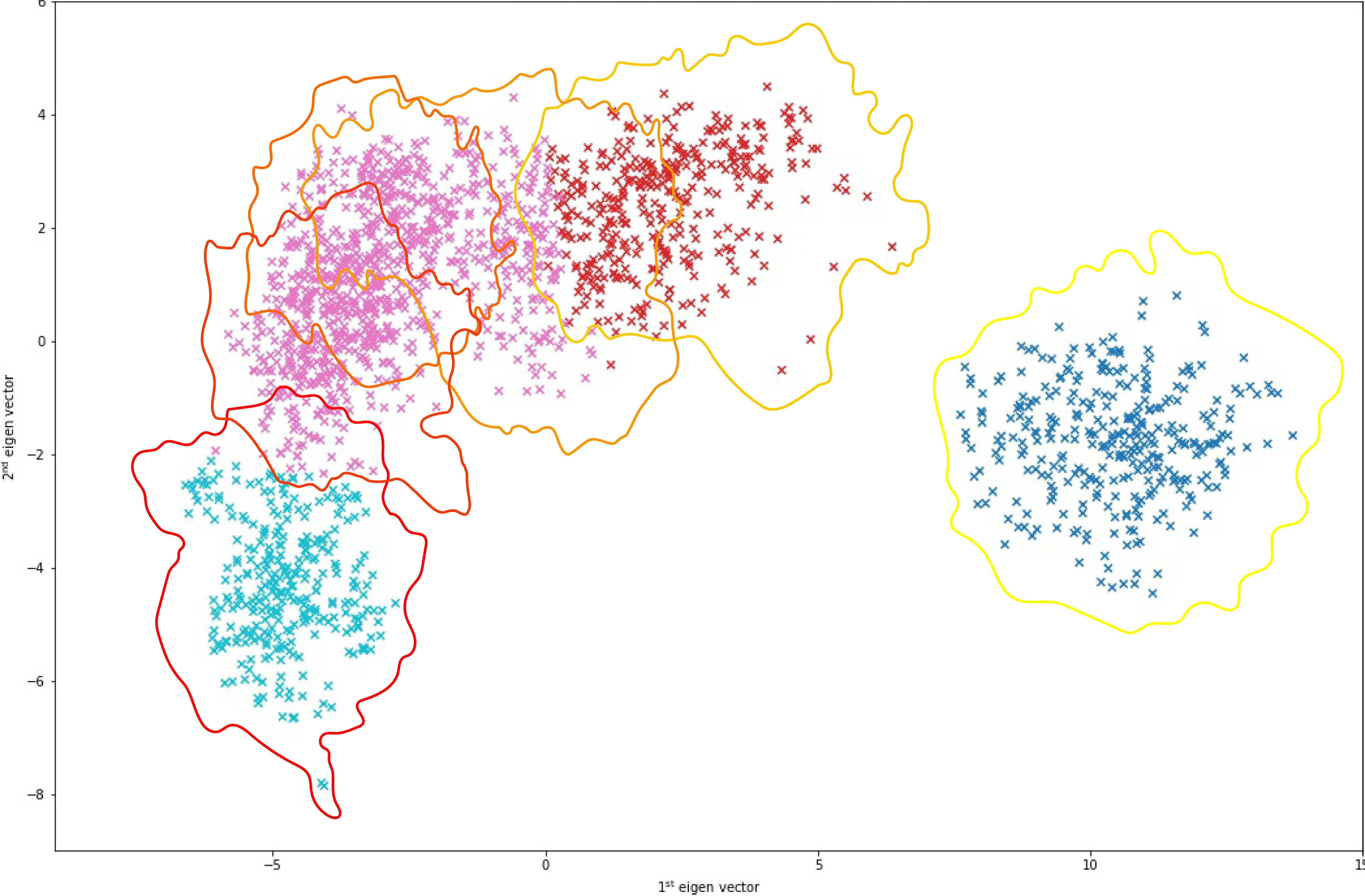}
  \caption{}
  \label{figure: lgv clusters overall trial}
\end{subfigure}

\medskip
\begin{subfigure}{0.33\textwidth}
  \captionsetup{justification=centering}
\includegraphics[width=\linewidth]{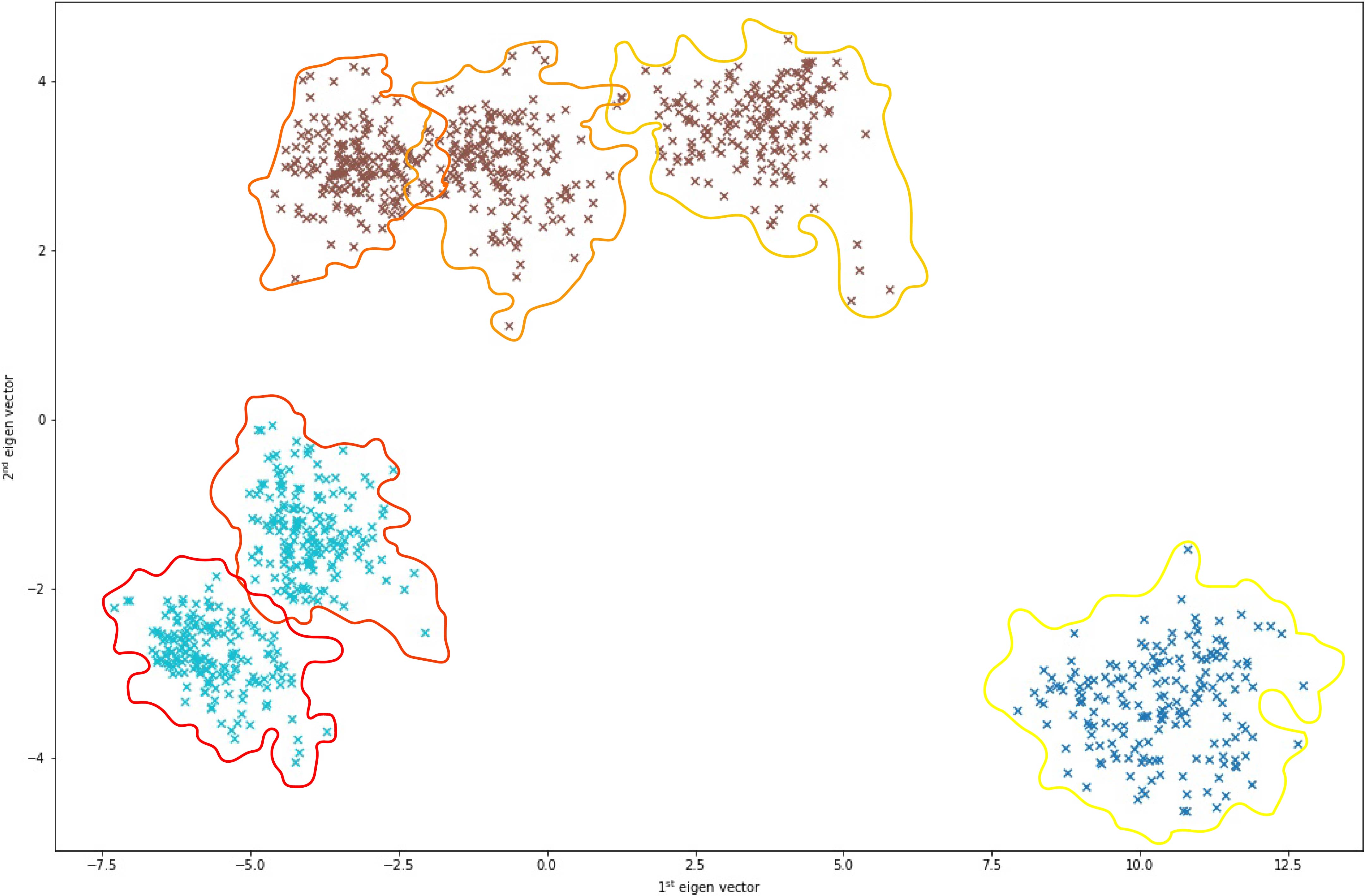}
  \caption{}
  \label{figure: lbw clusters trial 0}
\end{subfigure}
\begin{subfigure}{0.33\textwidth}
  \captionsetup{justification=centering}
  \includegraphics[width=\linewidth]{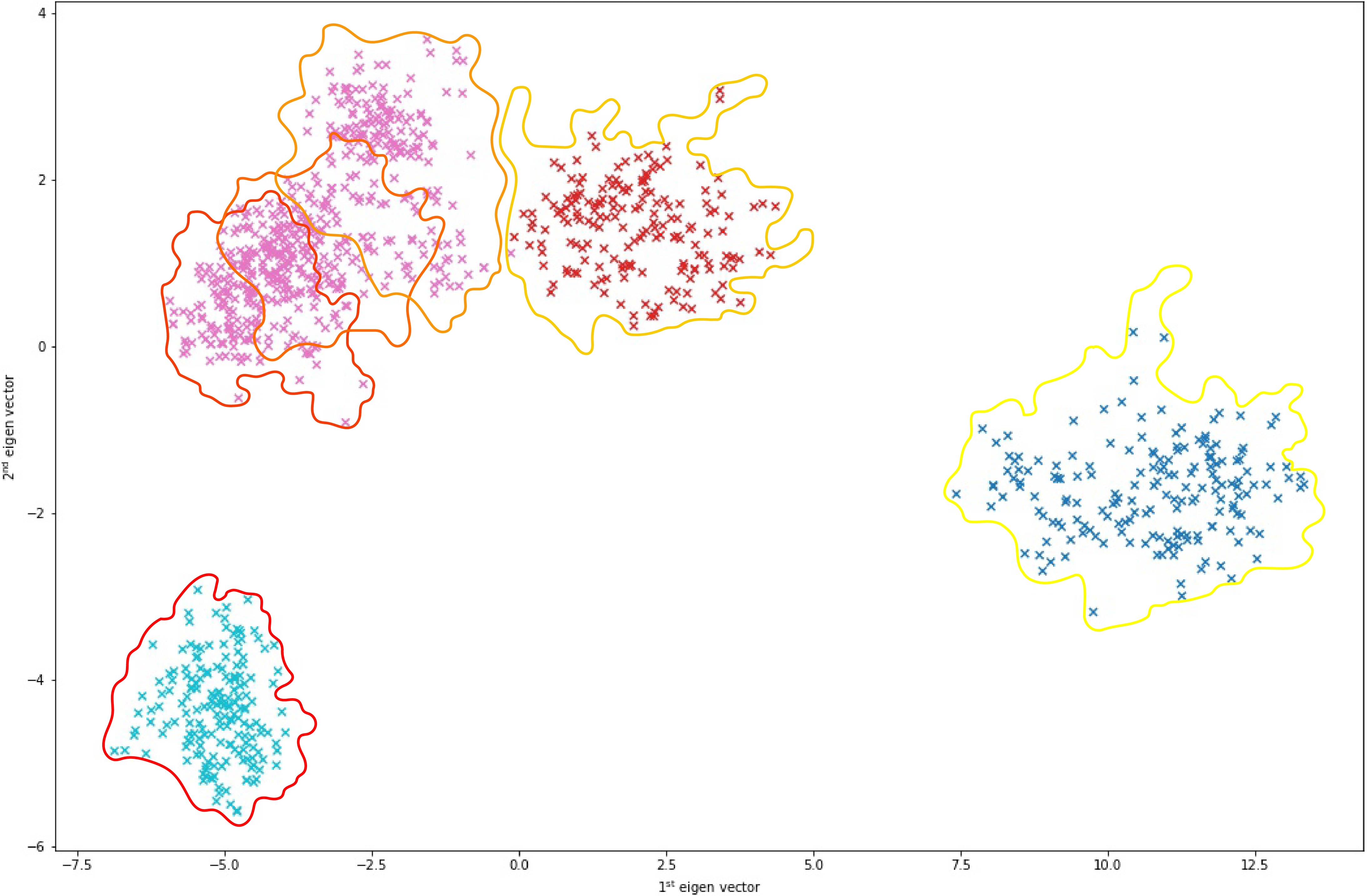}
  \caption{}
  \label{figure: lbw clusters trial 5}
\end{subfigure}
\begin{subfigure}{0.33\textwidth}
  \captionsetup{justification=centering}
  \includegraphics[width=\linewidth]{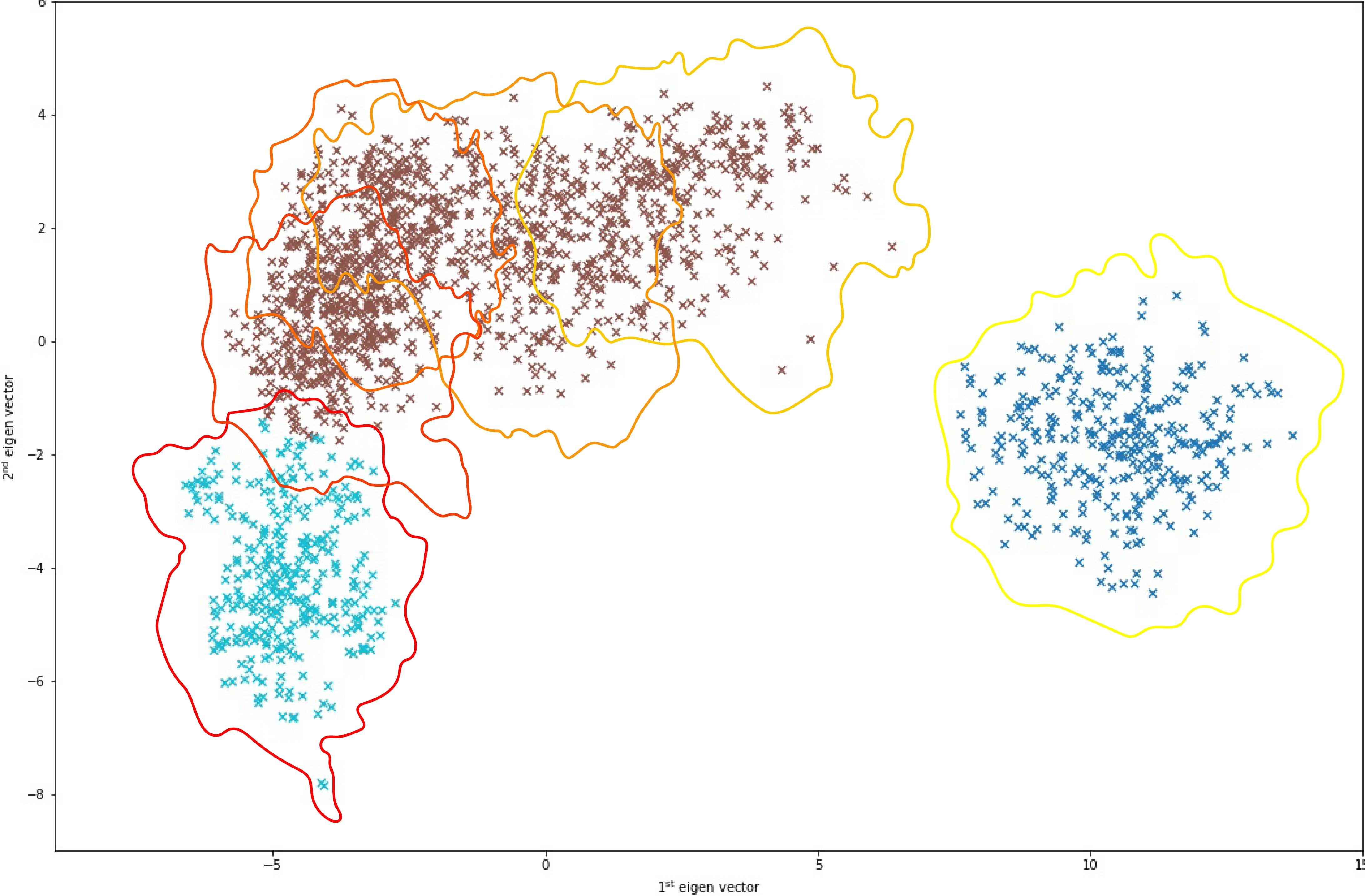}
  \caption{}
  \label{figure: lbw clusters overall trial}
\end{subfigure}

\medskip
\begin{subfigure}{\textwidth}
    \includegraphics[width=\textwidth]{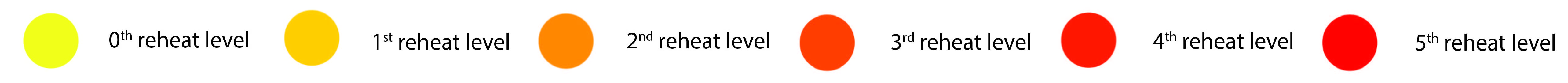}
    \captionsetup{justification=centering}
    \caption*{Legend for reheat cycle count classes}
\end{subfigure}
\caption{Clustering results of reheat cycle count classes under LGV algorithm for (a) trial $0$, (b) trial $5$, and (c) overall trials and under LBW algorithm for (d) trial $0$, (e) trial $5$, and (f) overall trials}
\label{figure: clustering results with spectral clustering}
\end{figure*}

In this section, the operation of the SC framework to distinguish noticeable changes in the oil as the reheat cycle count class is waxed exemplified for several trials. In Fig. \ref{figure: mode selection with sigma sweep curves}, the variation of the eigengaps with $\sigma$ as introduced in section \ref{section: discrimination of appreciable alterations} are plotted. Then, the corresponding clustering configurations are presented in Fig. \ref{figure: clustering results with spectral clustering}. Separately, the change in thiobarbituric acid reactive substances (TBARS) and total oxidation (TOTOX) were measured from a chemical analysis for the trials to check the results from the SC framework. The measured TBARS and TOTOX for the overall trials and two randomly selected trials (trial 0 and trial 5) out of the nine independent trials are given in Table \ref{table: chemical analysis results} as percentages according to \cite{Kamisah12}. In the incremental percentage values (given in parentheses in Table \ref{table: chemical analysis results}), a 20\% and 200\% increase in TBARS and TOTOX values respectively were observed with $\textit{p}<0.05$ implying a statistically significant result. Table \ref{table: critical reheat stages} presents the reheat cycle count classes with an appreciable change in the respective property (spectral, TBARS, and TOTOX).

% sigma sweep results under two algorithm

% clustering results under two algorithm

\begin{table*}[t]

\centering
\fontsize{1}{2}\selectfont
\tiny
\caption{Percentage increase in TBARS and TOTOX levels at each reheat cycle count class with reference to the pure oil properties. Incremental change at consecutive cycles are given within bracket. Significant changes $p < 0.05 $ are in bold typeface and underlined.}
\resizebox{\textwidth}{!}{
\begin{tabular}{c c c c c c c}
\hline
\\[-1.05em]
\hline
& \multirow{2}{*}{\begin{tabular}{c} Chemical\\[-0.8ex] Property \end{tabular} } &\multicolumn{5}{c}{Reheat cycle count class}\\
\cline{3-7}
\noalign{\vskip 1mm}
&  & 1 & 2 & 3 & 4 & 5 \\[0.3ex]
\hline
\noalign{\vskip 1mm}  
% Overall trials results
% TBARS results
\multirow{4}{*}{\rotatebox[origin=c]{90}{Overall trials}}&\multirow{2}{*}{TBARS} & 78.5~$\pm$~35 & 86.0~$\pm$~35 & 99.5~$\pm$~40 & 117.5~$\pm$~40 & 142.0~$\pm$~40\\
&& (\underline{\textbf{+~78.5}}) & (+~7.5) & (+~13.5) & (+~18.0) & (\underline{\textbf{+~24.5}}) \\[1.0ex]
% TOTOX results
&\multirow{2}{*}{TOTOX} & 240~$\pm$~40 & 395~$\pm$~50 & 530~$\pm$~80 & 820~$\pm$~50 & 925~$\pm$~105\\
&& (\underline{\textbf{+~240}}) & (+~155) & (+~135) & (\underline{\textbf{+~290}}) & (+~105) \\[1.0ex]
\hline
\noalign{\vskip 1mm} 
% Trial 0 results
% TBARS results
\multirow{4}{*}{\rotatebox[origin=c]{90}{Trial 0}}&\multirow{2}{*}{TBARS} & 105.5 & 114.5 & 122 & 161.5 & 161.5\\
&& (\underline{\textbf{+~105.5}}) & (+~9.0) & (+~7.5) & (\underline{\textbf{+~39.5}}) & (+~0) \\[1.0ex]
% TOTOX results
&\multirow{2}{*}{TOTOX} & 250 & 455 & 625 & 1005 & 1140\\
&& (\underline{\textbf{+~250}}) & (\underline{\textbf{+~205}}) & (+~170) & (\underline{\textbf{+~380}}) & (+~135) \\[1.0ex]
\hline
\noalign{\vskip 1mm} 
% Trial 5 results
% TBARS results
\multirow{4}{*}{\rotatebox[origin=c]{90}{Trial 5}}&\multirow{2}{*}{TBARS} & 38.5 & 58.0 & 63.0 & 69.0 & 108.5\\
&& (\underline{\textbf{+~38.5}}) & (+~19.5) & (+~5.0) & (+~6.0) & (\underline{\textbf{+~39.5}}) \\[1.0ex]
% TOTOX results
&\multirow{2}{*}{TOTOX} & 250 & 445 & 530 & 615 & 845\\
&& (\underline{\textbf{+~250}}) & (+~195) & (+~85) & (+~85) & (\underline{\textbf{+~230}}) \\[1.0ex]

\hline 
\\[-1.05em]
\hline
\end{tabular}}
\label{table: chemical analysis results}
\end{table*}

% Critical reheat stages results
\begin{table*}[h!]
\centering
\caption{Critical reheat cycle count classes for the trials according to the results from the SC framework and chemical analysis}
\resizebox{\textwidth}{!}{
\begin{tabular}{c c c c c c c c c c c}
\hline\hline
\multicolumn{2}{c}{\multirow{2}{*}{Method}} & \multicolumn{9}{c}{Sample}\\
\cline{3-11}
\noalign{\vskip 1mm}
&\multicolumn{1}{c}{} & Trial 0 & Trial 1 & Trial 2 & Trial 3 & Trial 4 & Trial 5 & Trial 6 & Trial 7 & Trial 8\\
\hline
\noalign{\vskip 1mm}
\multirow{2}{*}{\begin{tabular}{c} SC\\[-0.8ex] framework \end{tabular} } & LGV \smallskip&1,\,2,\,4 & 1,\,2,\,5 & 1,\,4,\,5 & 1,\,3,\,4 & 1,\,5 & 1,\,2,\,5 & 1,\,3,\,5 & 1,\,4,\,5 & 1,\,3\\
& LBW & 1,\,4 & 1,\,5 & 1,\,5 & 1,\,4 & 1,\,5 &  1,\,5 & 1,\,3 & 1,\,4 & 1,\,3\\
\multicolumn{2}{r}{} &\multicolumn{9}{c}{}\\
\multirow{2}{*}{\begin{tabular}{c} Chemical\\[-0.8ex] Analysis \end{tabular}  } & TBARS \smallskip & 1,\,4 & 1,\,5 & 1,\,5 & 1,\,4 & 1,\,5 &  1,\,5 & 1,\,3 & 1,\,4 & 1,\,3\\
& TOTOX & 1,\,2,\,4 & 1,\,2,\,5 & 1,\,5 & 1,\,4 & 1,\,5 &  1,\,5 & 1,\,3 & 1,\,4,\,5 & 1,\,3\\
\hline\hline

\end{tabular}}
\label{table: critical reheat stages}
\end{table*}

\section{Discussion}
\label{section: discussion}
This work proposed a novel application for MISs and their employment in food quality analysis. There is a conspicuous variance amongst the spectral properties of different reheat cycle count classes according to the false RGB representation given in Fig. \ref{figure: false rgb images of oil} which promotes the use of MSIs to detect and estimate the effects of reheating. Besides that, the insignificant differences between the RGB images of different reheat cycle count classes in Fig. \ref{figure: rgb images of oil} corroborate the applicability of transmittance MSIs over RGB photography with translucent specimens. Also, the mean Bhattacharyya distances in adjacent reheat cycle count classes in Fig. \ref{figure: rgb images of oil} for RGB photography are marginally separable compared to the values obtained with MSIs as provided in Fig. \ref{figure: false rgb images of oil}. In addition, the monotonic variation in Bhattacharyya distances observed for MSIs has not been replicated with RGB photographs. Furthermore, the variation observed in spectral properties with the reheat cycle count class has reemerged in Fig. \ref{figure: variation of bhattacharyya distances}. This result first validates the use of the particular statistical measure in estimating the number of reheating cycles but essentially intimates the usage of the Bhattacharyya distance in estimating the underlying chemical process.

The mean Bhattacharyya distance recorded for each reheat cycle count class monotonically increased (Fig. \ref{figure: variation of bhattacharyya distances}) with the reheat cycle count class, and this monotonic variation allowed to define decision boundaries that gradually increase as observed in Fig. \ref{figure: decision boundaries}. Nonetheless, there were overlapping distance intervals in Fig. \ref{figure: variation of bhattacharyya distances} which indicates the existence of coinciding spectral properties of adjacent reheat cycle count classes. As established previously, these characteristics of spectral properties are reflections of the chemical properties of the oil, and the resulted overlapping intervals implicate the divergent nature in the oil chemistry and the similarity in chemical properties of adjacent reheat cycle count classes.

Further, using a statistical measure has inadvertently allowed the classifier to use the sample properties in estimating the reheat cycle count class analogous to a chemical analysis. Whereas, had the classifier been trained on spectral properties of individual pixels, the estimations of the pixels of a sample will inherit the distribution observed in the signatures. Therefore, the distribution of the classification results has to be considered to estimate the reheat cycle count class of the sample, which is antithetical to the use of individual spectral properties over sample properties. In addition, the application of a statistical measure offers flexibility in selecting a preferable sample size from the MSIs of a given oil sample. Moreover, the use of a statistical measure could be construed as observing a superpixel that inherits the spectral properties of that sample. Hence, the proposed method could even be used with low image resolutions because low-resolution images will record the cumulative effect --- similar to using a statistical measure --- of spectral signatures under observations. Besides, high-resolution images inherently perform superior as these images could contain more delicate information of the spectral properties. 

The decision boundaries given in Fig. \ref{figure: decision boundaries} have distinctly separated the pure oil samples from heated oil, indicating the spectral properties are drastically changed. Also, this distinction has been recorded in confusion matrices for both train and test data with a classification accuracy of 100\%. Overall, the SVM classifier has recorded an overall accuracy of 86.84\% and 83.34\% for train and test data. Since the classifier can differentiate pure oil from heated oil, it is apt to reconsider the classifier's performance with heated oil only. With only heated oil samples, the classifier can estimate the reheat cycle count class with accuracy levels of 84.21\% and 80.00\% for train and test data, respectively. 

The SC framework was introduced to detect the reheat cycle count classes where a marked change in the spectral properties happened. In both illustrated sigma-sweep curves in Fig. \ref{figure: lgv mode selection trial 0} and \ref{figure: lgv mode selection trial 5}, the dominant clustering mode has been recorded as four under the LGV algorithm. However, the corresponding clustering configuration of the two trials is disparate according to Fig. \ref{figure: lgv clusters trial 0} and \ref{figure: lgv clusters trial 5}. According to Fig. \ref{figure: lgv clusters trial 0}, the framework has chosen the first, second, and fourth reheat cycle count classes as the stages where a drastic alteration has occurred in the oil sample in trial 0. When compared with the chemical results in Table \ref{table: chemical analysis results} for trial 0, a similar pattern is noticeable for the TOTOX level, whereas, with TBARS value, a suitable separation would be the first, and fourth as the critical reheat cycle count classes. In fact, in Fig. \ref{figure: lgv mode selection trial 0}, the maximum gap value of the third and fourth eigengap is similar, and the prominent mode of 4 has been chosen with a slight margin. Furthermore, the SC framework has chosen first, second, and fifth as the reheat cycle count classes of interest for trial 5, and the separation could be observed in both the TBARS and TOTOX values from the chemical analysis. However, similar to trial 0, the optimal separation could be redressed to three prominent clusters with the first and fifth cycles as critical.

The variation in sigma-sweep curves, for both the trials as $\sigma$ is extremely increased, complements the above correction for the prominent mode, as the value of the third eigengap supersedes that of the fourth eigengap. Besides that, the commonality in both these cases is that the eigengap (mode) that sustained the gap value over an extensive range of $\sigma$ is the same and is three as presented in Fig. \ref{figure: lbw mode selection trial 0} and \ref{figure: lbw mode selection trial 5}. The clustering results with the LBW algorithm are presented in Fig. \ref{figure: lbw clusters trial 0} and \ref{figure: lbw clusters trial 5}. When the LGV algorithm and the LBW algorithm are juxtaposed, each method has its merits and demerits. For example, the LGV algorithm tends to overestimate the number of critical reheat stages but guarantees the maximal closeness of the spectral properties of the reheat stages that are grouped. Whereas the LBW algorithm is more robust in operation and is immune to noise in spectral properties, yet could underestimate the number of critical stages as the algorithm seeks for maximally and equally separated clusters. Further, both the LGV and LBW methods grouped reheat cycle count classes with slight alterations in spectral properties while dividing at reheat cycle count classes with significant changes. Thereby, the inter-class boundaries represented appreciable changes in composition, which indicates significant chemical property changes that could be a potential health risk.

The performance of the LGV algorithm on an ensemble of reheated oils are illustrated in Fig. \ref{figure: lgv mode selection overall trials} and Fig. \ref{figure: lgv clusters overall trial} along with the chemical analysis results in Table \ref{table: chemical analysis results}. The prominent clustering mode given by the framework is four since the fourth eigengap is the largest. However, the clustering configuration of the mixture of an oil sample is inconsistent with the results of the chemical analysis even though the individual trials complied with the framework results. This inconsistency could have been influenced by the spectral similarity of adjacent reheat cycle count classes of different trials. Hence, the SC framework has recognized the first, second, and fifth reheat cycle count class as critical, while the chemical analysis ruled first and fourth cycles and first and fifth cycles as critical for TOTOX and TBARS values, respectively. Nonetheless, the LBW algorithm has managed to replicate the results (Fig. \ref{figure: lbw mode selection overall trials} and Fig. \ref{figure: lbw clusters overall trial}) for TBARS values of the mixture and this agreement of results powerfully portrays the association between the spectral properties and TBARS values. Despite the hodgepodge of spectral signatures from different trials, the framework has managed to group signatures from the same reheat cycle count class and minimize misclassifications.

In this paper, the experimental procedure mimics edible oils for frying in foodservice establishments. The proposed application could be used to estimate the reheat cycle count class of coconut oil and to visualize the harm to the chemical properties of the oil as the reheat cycle count class is increased. The food authorities could use the application to check the use of reheated oil at food establishments. Also, the vendors could use the MIS as visual assistance for safe reheating and detect significant changes to the oil, which could cause the oil to be deemed unfit for consumption.

\section{Conclusion}
\label{section: conclusion}

In this paper, a novel application was proposed for MISs to estimate reheat cycle count class and discrimination of appreciable alterations in the chemical and thermophysical properties under repetitive heating for frying oil, with coconut oil as the case study. The proposed work introduced the transmittance configuration of the MIS to acquire images of translucent liquid specimens instead of the conventional reflectance configuration. It was observed that the reheating of oil is another oil degradation process like adulteration, and MSIs prevail over RGB images in the detection of these corruptions. The spectral properties were proven to be sensitive to the reheat cycle count class of the oil with the incorporation of Bhattacharyya distance. The SVM classifier had an accuracy of 80.00\,\%, 83.34\,\%, and 100\,\% for estimating reheat cycle count classes excluding pure oil, including pure oil, and separation of pure oil and heated oil, respectively, for test data. Furthermore, a novel algorithm was discussed to select the prominent mode and dominant $\sigma$ for the SC framework suggested for detecting critical reheat cycle count classes using spectral properties.
The proposed LBW algorithm identified clusters of reheat cycle count classes that are dramatically separated in terms of spectral properties due to the significant changes in the chemical properties. The identification was performed using a modified version of SC to group reheat cycle count classes that were only marginally separated. The modified version resulted in the formation of umbrella clusters where the separation was significant in the feature space. The cluster formation was shown to be in line with the results of the chemical validation performed subsequently. At the same time, the LGV algorithm promoted the connectivity of intra-class spectral signatures in each umbrella cluster more than the separation between those umbrella clusters. Hence, while the LGV algorithm managed to reproduce the results from the LBW algorithm, it was able to identify `reheat cycle count classes' with minor alterations as well.
Furthermore, it was evident that for proper operation of the framework, it is necessary to analyze different oil samples separately, similar to alternative analytical techniques, and should not be amalgamated. The chemical analysis yielded that the TBARS and TOTOX values of oil significantly changed with the reheat cycle count classes, and the results of the SC framework were deemed to coincide with the formerly mentioned chemical property alterations. Food authorities and foodservice establishments could use the proposed application for adherence to health and safety protocol regarding safe reheating of oil. However, the case study used in this work was limited to coconut oil and potato chips as fryable food. Hence, we wish to expand the proposed application and frameworks to other oil types such as palm and soybean and other fryable food types. Also, the MIS could be extended to measure the change in TBARS and TOTOX values as there was an appreciable variation (\textit{p $< 0.05$}) in TBARS and TOTOX readings as the reheat cycle count class was incremented.

\begin{appendices}

\section{}
\label{appendix: A}

\setcounter{table}{0}
\renewcommand{\thetable}{A\arabic{table}}

The reheat count class classification accuracies of SVM classifier were compared with different classifiers: Neural networks, Nearest neighbor classifier, Nearest centroid classifier, Gaussian process classifier, and Random Forrest. The comparison results are given in Table \ref{table: classifier comparison}.

\begin{table*}[h]
\centering
\caption{Classification accuracy for reheat count class classification with different classifiers}
\label{table: classifier comparison}
\Large
\resizebox{\textwidth}{!}{
\begin{tabular}{c c c c c c c}

\hline\hline

{\multirow{3}{*}{\begin{tabular}{c} Reheat count\\[-0.8ex] class \end{tabular}}} & \multicolumn{6}{c}{Classifier}\\\cline{2-7}
&{\multirow{2}{*}{\begin{tabular}{c} Support Vector\\[-0.8ex] Machine \end{tabular}}} &{\multirow{1}{*}{\begin{tabular}{c} Neural networks\\[-0.8ex]  \end{tabular}}}  &{\multirow{2}{*}{\begin{tabular}{c} Nearest Neighbor\\[-0.8ex] Classifier \end{tabular}}}  &{\multirow{2}{*}{\begin{tabular}{c} Nearest Centroid\\[-0.8ex] Classifier \end{tabular}}} &{\multirow{2}{*}{\begin{tabular}{c} Gaussian Process\\[-0.8ex] Classifier \end{tabular}}} &{\multirow{1}{*}{\begin{tabular}{c} Random Forrest\\[-0.8ex]  \end{tabular}}}\\
\noalign{\vskip 5mm}
\hline 
\setlength\extrarowheight{5pt}
0	&1.000	&1.000	&1.000	&1.000	&1.000	&1.000\\
1	&0.909	&-	&0.886	&1.000	&1.000	&0.800\\
2	&0.682	&1.000	&0.486	&0.514	&0.457	&0.600\\
3	&0.864	&-	&0.486	&0.800	&0.714	&0.429\\
4	&0.727	&-	&0.314	&0.600	&0.229	&0.257\\
5	&0.818	&1.000	&0.029	&-	&-	&0.057\\
\hline
overall	&0.8334	&0.5000	&0.5335	&0.6524	&0.5664	&0.5238\\
\hline\hline
\end{tabular}}
\end{table*}

\section{}
\label{appendix: B}

\setcounter{table}{0}
\renewcommand{\thetable}{B\arabic{table}}

\begin{table*}[b]
\centering
\caption{Parametric sweep of $\gamma$ and cost for the SVM classifier. Highest accuracy is in bold typeface.}
\label{table: parametric sweep for the classifier}

\resizebox{\textwidth}{!}{
\begin{tabular}{c l c c c c c c c c c c}

\hline\hline
\multicolumn{2}{c}{\multirow{2}{*}{\begin{tabular}{c} Classification\\[-0.8ex] Accuracy \end{tabular}}} & \multicolumn{10}{c}{$\gamma$}\\\cline{3-12}
\noalign{\vskip 1mm}
& &0.5 &0.6 &0.7 &0.8 &0.9 &1.0 &1.1 &1.2 &1.3 &1.4 \\
\hline % cline{3-12}
\noalign{\vskip 1mm}
\multirow{8}{1mm}{\rotatebox[origin=c]{90}{\hspace{4mm}Cost}}
&\hspace{5mm}$10^{-3}$  &0.7997 &0.8206 &0.8206 &0.8218	&0.8206	&0.8229	&0.8276	&0.8299	&0.8334	&0.8334 \\

&\hspace{5mm}$10^{-2}$ &0.7997 &0.8206 &0.8206 &0.8218	&0.8206	&0.8229	&0.8276	&0.8299	&0.8334	&0.8334 \\

&\hspace{5mm}$10^{-1}$ &0.8113 &0.8253 &0.8299 &0.8287 &0.8276	&0.8311	&0.8311	&0.8345	&0.8334	&0.8369 \\

&\hspace{5mm}$10^{0}$ &0.8160	&0.8299	&0.8334	&0.8345	&0.8392	&0.8369	&0.8380	&0.8380	&0.8369	&0.8357  \\

&\hspace{5mm}$10^{1}$ &0.8183	&0.8345	&0.8369	&0.8369	&0.8369	&0.8334	&0.8369	&0.8369	&0.8415	&0.8427 \\

&\hspace{5mm}$10^{2}$ &0.8241	&0.8334	&0.8380	&0.8369	&0.8345	&0.8369	&0.8427	&\textbf{0.8684} &0.8438	&0.8450  \\

&\hspace{5mm}$10^{3}$ &0.8264	&0.8369	&0.8369	&0.8345	&0.8369	&0.8438	&0.845	&0.8427	&0.8462	&0.8415 \\

&\hspace{5mm}$10^{4}$ &0.8253	&0.838	&0.8369	&0.8369	&0.8438	&0.8427	&0.8438	&0.8438	&0.8415	&0.8462 \\

\hline\hline
\end{tabular}}
\end{table*}

The classification accuracy from the parametric sweep are given in Table \ref{table: parametric sweep for the classifier} for the selection of optimal $\gamma$ and cost.

\end{appendices}

\section*{Acknowledgment}
The authors thank Ms. E.G.T.S. Wijethunga, Department of Food Science and Technology, University of Peradeniya, Sri Lanka for assisting the chemical analysis of this work. Also, the authors acknowledge Silver Mills Groups, Meerigama for supplying coconut oil.

%\clearpage
%\doublespacing
\bibliographystyle{IEEEtran}
\bibliography{IEEEabrv,bibliography}

%\clearpage

\singlespacing
\clearpage

\begin{IEEEbiography}[{\includegraphics[width=1in,height=1.25in,clip]{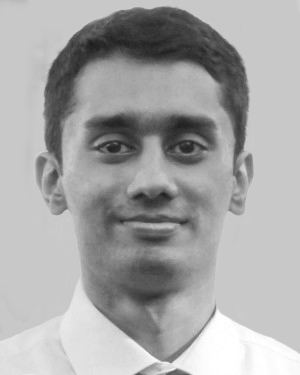}}]{D.Y.L. Ranasinghe} received his B.Sc. Engineering degree in Electrical and Electronic Engineering from the University of Peradeniya, Sri Lanka, in 2020. Immediately after, he joined the School of Engineering, Sri Lanka Technological Campus, Padukka, Sri Lanka as a Research Assistant. He is currently working as a research assistance at University of Peradeniya under a research grant from the International Development Research Centre (IDRC), Canada. His research interests include hyperspectral and multispectral imaging, remote sensing, signal and image processing, and deep learning. He has numerous publications in IEEE conferences.     
\end{IEEEbiography}

\vskip -30pt plus -1fil

\begin{IEEEbiography}[{\includegraphics[width=1in,height=1.25in,clip,keepaspectratio]{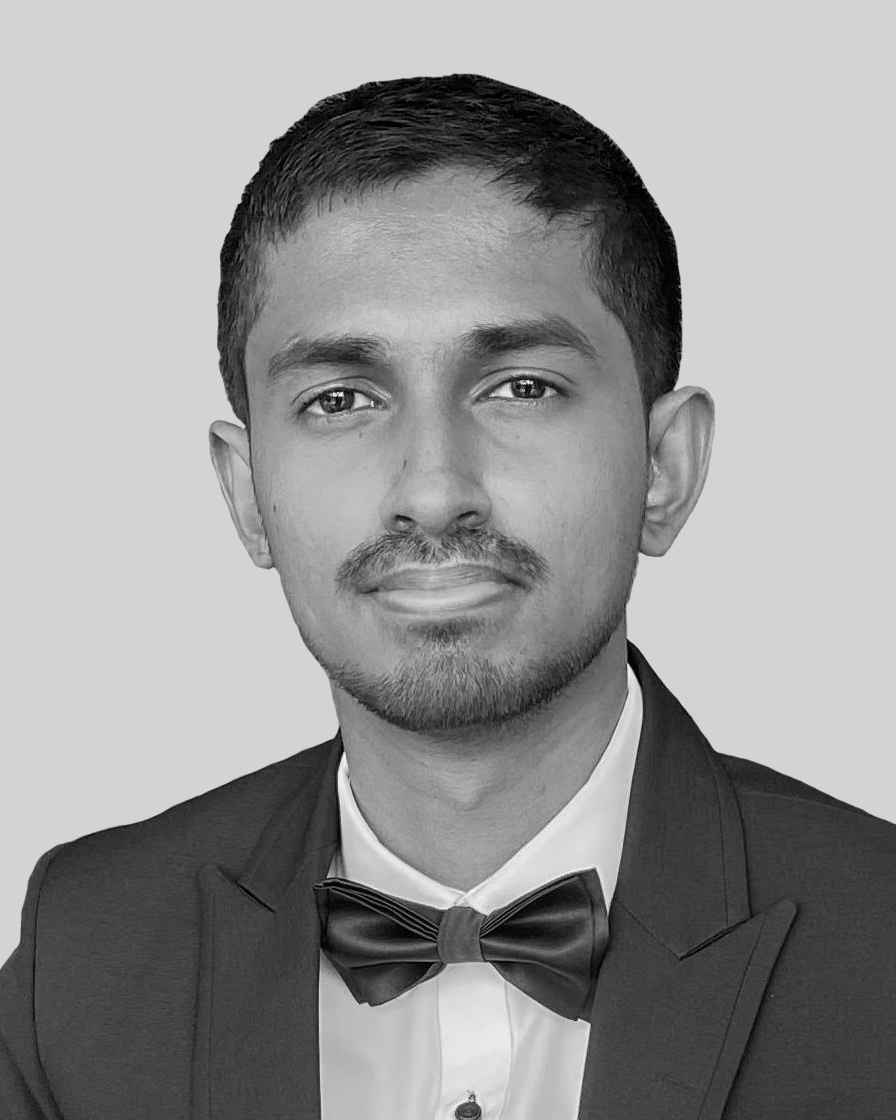}}]{H.M.H.K. Weerasooriya} obtained his degree in Electrical and Electronic Engineering with first-class honours and he currently works as an Instructor in the Department of Electronic and Electrical Engineering. Currently, he is involved in the researches on hyperspectral imaging for remote sensing and agriculture applications, and he has numerous publications in IEEE conferences. His research interests include image processing, signal processing, communication, machine learning and deep learning.
\end{IEEEbiography}
\vskip -30pt plus -1fil

\begin{IEEEbiography}[{\includegraphics[width=1in,height=1.25in,clip]{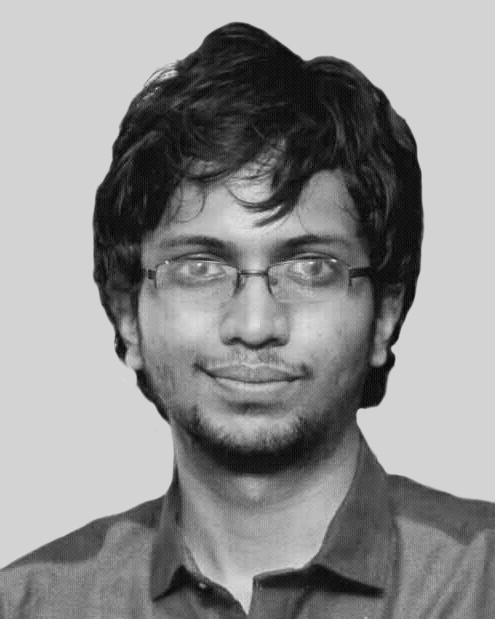}}]{S. Herath} received his B.Sc. Engineering degree in Electrical and Electronic Engineering from the University of Peradeniya, Sri Lanka, in 2020. Immediately after, he joined the Department of Engineering Mathematics, University of Peradeniya as a Teaching Instructor. He is currently a graduate student at the Department of Electrical and Computer Engineering, University of Maryland, USA. His research interests include computer vision, image and signal processing, pattern recognition, blind source separation, and machine learning. He has numerous publications in IEEE conferences.      
\end{IEEEbiography}

\vskip -30pt plus -1fil

\begin{IEEEbiography}[{\includegraphics[width=1in,height=1.25in,clip]{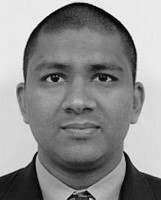}}]{M.P.B. Ekanayake} (Senior Member, IEEE) received his B.Sc. Engineering degree in Electrical and Electronic Engineering from University of Peradeniya, Sri Lanka, in 2006, and Ph.D. from Texas Tech University in 2011. Currently, he is attached to the University of Peradeniya as a Senior Lecturer.

His current research interests include applications of signal processing and system modeling in remote sensing, hyperspectral imaging, and smart grid. He is a Senior Member of the IEEE. He is a recipient of the Sri Lanka President's Award for Scientific Publications in 2018 and 2019. He has obtained several grants through the National Science Foundation (NSF) for research projects. His previous works have been published in IEEE-TGRS and several other IEEE-GRSS conferences including WHISPERS and IGARSS. He also has multiple publications in many IEEE transactions, Elsevier and IET journals and has been awarded several best paper awards in international conferences.
	
\end{IEEEbiography}

\vskip -30pt plus -1fil

\begin{IEEEbiography}[{\includegraphics[width=1in,height=1.25in,clip]{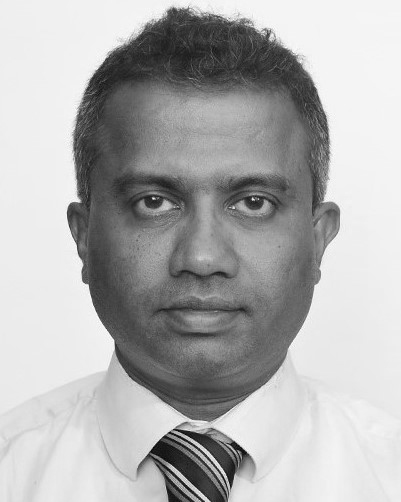}}]{H.M.V.R. Herath} (Senior Member, IEEE) received the B.Sc.Eng. degree in electrical and electronic engineering with 1st class honours from the University of Peradeniya, Peradeniya, Sri Lanka, in 1998, M.Sc. degree in electrical and computer engineering with the award of academic merit from the University of Miami, USA in 2002, and Ph.D. degree in electrical engineering from the University of Paderborn, United States in 2009. In 2009, he joined the Department of Electrical and Electronic Engineering, University of Peradeniya, as a Senior Lecturer.

His current research interests include hyperspectral imaging for remote sensing, multispectral imaging for food quality assessment, Coherent optical communications and integrated electronics. Dr. Herath was a member of one of the teams that for the first time successfully demonstrated coherent optical transmission with QPSK and polarization multiplexing. He is a member of the Institution of Engineers, Sri Lanka and The Optical Society. He is a Senior Member of the IEEE. He was the General Chair of the IEEE International Conference on Industrial and Information Systems (ICIIS) 2013 held in Kandy, Sri Lanka. His previous works have been published in IEEE-TGRS and several other IEEE-GRSS conferences including WHISPERS and IGARSS. He received the paper award in the ICTer 2017 conference held in Colombo Sri Lanka. Dr. Herath is a recipient of Sri Lanka President's Award for scientific research in 2013.
\end{IEEEbiography}

\vskip -30pt plus -1fil

\begin{IEEEbiography}[{\includegraphics[width=1in,height=1.25in,clip]{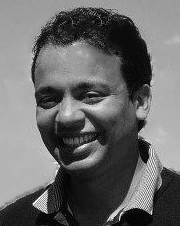}}]{G.M.R.I. Godaliyadda} (Senior Member, IEEE) obtained his B.Sc. Engineering degree in Electrical and Electronic Engineering from the University of Peradeniya, Sri Lanka, in 2005, and Ph.D. from the National University of Singapore in 2011. Currently, he is attached to the University of Peradeniya, Faculty of Engineering, Department of Electrical and Electronic Engineering as a Senior Lecturer.

His current research interests include image and signal processing, pattern recognition, computer vision, machine learning, smart grid, bio-medical and remote sensing applications and algorithms. He is a Senior Member of the IEEE. He is a recipient of the Sri Lanka President's Award for Scientific Publications for 2018 and 2019. He is the recipient of multiple grants through the National Science Foundation (NSF) for research activities. His previous works have been published in IEEE-TGRS and several other IEEE-GRSS conferences including WHISPERS and IGARSS. He also has numerous publications in many other IEEE transactions, Elsevier and IET journals and is the recipient of multiple best paper awards from international conferences for his work.
\end{IEEEbiography}

\vskip -30pt plus -1fil

\begin{IEEEbiography}[{\includegraphics[width=1in,height=1.25in,clip]{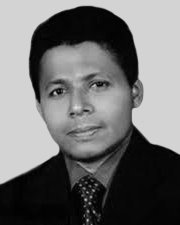}}]{Terrence Madujith} obtained his B.Sc. (Agriculture) from University of Peradeniya in 1994 and served as a food technologist at Lanka Canneries Ltd for a short period. Subsequently, he joined the Department of Food Science and Technology as a lecturer and proceeded to Canada to pursue higher studies. He received M.Sc. and Ph.D. in Food Science from Memorial University of Newfoundland, Canada. He was promoted to Professor in Food Science and Technology in 2015 and subsequently to the Chair Professor of Food Science and technology in 2019. He also serves as the Director of the Research Council of University of Peradeniya. His research interests include lipid chemistry, food safety and food microbiology.
\end{IEEEbiography}
\EOD
\end{document}